\documentclass[useAMS,usenatbib]{mn2e}

\usepackage{amsmath}
\usepackage{amssymb}
\usepackage[dvipsnames]{xcolor}
\usepackage{graphicx}
\usepackage{graphics}
\usepackage{hyperref}
\usepackage{multirow}
\usepackage{multicol}
\usepackage{comment}
\usepackage{ulem}
\usepackage[autostyle]{csquotes}

\hypersetup{
     colorlinks   = true,
     citecolor    = blue
}

\def\apj{{ ApJ}}
\def\apjl{{ApJL}}
\def\apjs{{ ApJS}}

\def\aap{{ A\&A}}

\def\aj{{ AJ}}
\def\mnras{{ MNRAS}}
\def\araa{{ ARA\&A}}

\def\nat {{ Nature}}

\def\ssr{{ Space Sci. Rev.}}

\def\prd{{ Phys. Rev. D}}
\def\prl{{ Phys. Rev. L}}

\def\mr{\mathrm}
\def\d{\mr{d}}
\def\b{\mathbf}

\def\mc{\mathcal}
\def\msun{M_{\rm \odot}}
\def\eps{\mc{E}}

\def\rp{r_{\rm p}}

\def\mns{m_{\rm ns}}
\def\vk{v_{\rm k}}
\def\bvk{\b{v}_{\rm k}}
\def\hv{\hat{\b{v}}}
\def\hx{\hat{\b{x}}}
\def\hz{\hat{\b{z}}}

\def\cot{\mr{cot}\,}

\newcommand{\lara}[1]{{\langle#1\rangle}}

\newcommand{\myemail}{wenbinlu@caltech.edu}

\title[GW190814]{On the formation of GW190814}
\author[Lu, Beniamini, Bonnerot]
  {Wenbin Lu$^1$\thanks{\myemail}, Paz Beniamini$^1$, and Cl{\'e}ment Bonnerot$^1$\\
  $^1$Theoretical Astrophysics, Mail Code 350-17, Caltech, Pasadena, CA 91125, USA\\}

\begin{document}
\label{firstpage}
\maketitle

\begin{abstract}
The LIGO-Virgo collaboration recently reported a puzzling event, GW190814, with component masses of 23 and $2.6\msun$.
Motivated by the relatively small rate of such a coalescence (1--$23\rm\, Gpc^{-3}\, yr^{-1}$) and the fact that the mass of the secondary is close to the total mass of known binary neutron star (bNS) systems, we propose that GW190814 was a second-generation merger from a hierarchical triple system, i.e., the remnant from the bNS coalescence was able to merge again with the $23\msun$ black hole (BH) tertiary. We show that this occurs at a sufficiently high probability provided that the semimajor axis of the outer orbit is less than a few AU at the time of bNS coalescence. It remains to be explored whether the conditions for the formation of such tight triple systems are commonly realized in the Universe, especially in low metallicity ($\lesssim 0.1Z_\odot$) environments. Our model provides a number of predictions. (1) The spin of the secondary in GW190814-like systems is 0.6--0.7. (2) The component mass distribution from a large sample of LIGO sources should have a narrow peak between $2.5$ and $\sim$3.5$\msun$, whereas the range between $\sim$3.5 and $\sim$5$\msun$ stays empty (provided that stellar evolution does not generate such BHs in the ``mass gap"). (3) About 90\% (10\%) of GW190814-like events have an eccentricity of $e\gtrsim 2\times10^{-3}$ ($\gtrsim 0.1$) near gravitational wave frequency of $10\,$mHz. (4) A significant fraction ($\gtrsim10\%$) of bNS mergers should have signatures of a massive tertiary at a distance of a few AU in the gravitational waveform. (5) There are $10^5$ undetected radio-quiet bNS systems with a massive BH tertiary in the Milky Way.
\end{abstract}


\section{Introduction}
The formation history of the gravitational wave (GW) sources detected by LIGO-Virgo collaboration is one of the major unsolved questions in astrophysics today. The main difficulty is because LIGO/Virgo detectors are only sensitive to a tiny fraction --- the very last few seconds of GW-driven evolution --- of the entire history of the source.
Observations typically do not directly provide a unique solution of their formation pathways, although statistical inference from a sufficiently large sample is still possible \citep{LVC19_bbh_population}.

The recent event GW190814 had component masses of 23 and 2.6$\,M_\odot$, each measured accurately to within $\lesssim 5\%$ \citep{abbott20_GW190814}. The effective spin\footnote{The effective spin describes the mass-weighted sum of the spins projected along the orbital angular momentum $\chi_{\rm eff} = (m_1\chi_1+m_2\chi_2)/(m_1+m_2)$, where $\chi_i=\boldsymbol{\chi}_i\cdot\hat{\b{L}}$, $\boldsymbol{\chi}_i$ and $m_i$ are the dimensionless spin angular momentum and mass of object $i(=1, 2)$ respectively, and $\hat{\b{L}}$ is the unit vector along the orbital angular momentum of the binary. } was $\chi_{\rm eff}=0\pm 0.06$ and the dimensionless spin of the primary is constrained to be less than 0.07, but the spin of the significantly less massive secondary is essentially unconstrained due to low signal-to-noise ratio.  The nature of the low-mass component is debated between an unprecedented massive neutron star (NS) or an extremely low-mass black hole (BH).

Associating the $2.6\msun$ component with a NS is challenging in light of the binary neutron star (bNS) merger event GW170817 \citep{LVC_GW170817_parameters}.
This event left over a compact remnant of mass\footnote{The system lost about $0.05\msun$ in baryonic ejecta as inferred from the kilonova emission \citep[e.g.,][and references therein]{metzger17_kilonova_review}, $0.05$--$0.1\msun$ in gravitational waves, possibly $\lesssim 0.1\msun$ in neutrinos if the remnant is sufficiently long-lived to undergo neutrino cooling.} $\approx 2.6\msun$, which is comparable to that of the secondary in GW190814. The electromagnetic counterpart of GW170817 strongly disfavored a long-lived NS remnant, which would otherwise overproduce the X-ray flux at late time and inject too much energy in the relativistic ejecta to be consistent with the afterglow data \citep[e.g.,][]{Granot2017,pooley18_GW170817_BH, margutti18_GW170817_afterglow, xie18_jet_simulation, margalit19_multimessenger_matrix, makhathini20_GW170817afterglow_data, salafia20_jet_energy}. Based on the assumption that GW170817 made a BH, many authors have concluded that the maximum mass of a non-rotating NS is $M_{\rm TOV}\lesssim 2.3\msun$ \citep{margalit17_Mmax, rezzolla18_Mmax, shibata19_Mmax_from_GW170817} \citep[but see][for a discussion of the less likely case that GW170817 did not make a BH]{ai20_mtov}. Although rapid uniform rotation can support mass up to about $1.2M_{\rm TOV}$ \citep{rezzolla18_Mmax}, the low-mass component in GW190814 is most likely spinning slowly without significant centrifugal support (a NS near break-up rotation should have spun down during GW inspiral) and hence it is most likely a BH.

Direct formation of such low-mass BHs in the so-called ``mass gap" from core-collapse is disfavored by the Galactic BH mass distribution  \citep{ozel10_BH_mass_gap, farr11_BH_mass_distribution, ozel12_NS_birth_mass} and from current understanding of stellar evolution \citep{fryer12_BH_masses, zevin20_mass_gap_BH_from_stellar_evolution}. In fact, there is no confirmed case of such low-mass BH, although one candidate has recently been reported \citep[][but the inferred mass of this candidate object is only marginally consistent with the secondary of GW190814]{thompson19_low_mass_BH}. The mass of the secondary in GW190814 is, however, very similar to the typical total mass of Galactic bNS systems, which have accurately measured total masses in the range 2.5 to $2.9\msun$ \citep[][and references therein]{farrow19_bNS_mass_distribution}. The simplest explanation is therefore that the secondary in GW190814 is itself the product of a bNS merger.

The $23\msun$ component is also special in that massive stars near solar metallicity ($Z_\odot$) typically do not make such heavy BHs, as demonstrated by many stellar evolution studies \citep[e.g.,][]{spera15_BH_masses, belczynski16_lowZ, giacobbo18_bBH_masses, woosley19_He_star_evolution}. The final BH mass is determined by the metallicity-dependent mass loss from stellar winds and the ejecta mass during the supernova explosion. The formation of BHs with masses larger than $20\msun$ is expected to be efficient only at $0.1Z_\odot$ or less \citep{belczynski16_lowZ}. Empirically, from the high-mass X-ray binary M33 X-7, which consists of a BH mass of $15.6\pm 1.5\msun$ and a massive companion star of $70\pm 7\msun$ \citep{Orosz07_16Msun_BH}, we know that even stars with initial mass above $\sim70\msun$ do not make BHs of more than $20\msun$\footnote{The BH in the binary is most likely the outcome of the initially more massive star in the binary, which therefore also evolved faster.}. Thus, we expect the generation of GW180914-like systems to be rare\footnote{A possible exception is that stellar mass BHs can grow substantially in mass in gas-rich environment near active galactic nuclei \citep{yang19_bbh_AGN}.} in the local Universe. These systems are most likely to have been formed many Gyrs ago when most galaxies were less metal enriched. We also note that the multiplicity fraction for short period systems increases towards lower metallicity \citep{gao14_metallicity_binarity, yuan15_metallicity_binary_fraction, badenes18_metallicity_multiplicity, moe19_multiplicity}, although it is already very high for the most massive stars at solar metallicity, with an average number of $\sim$2 companions per central object \citep[][]{sana12_binary_interaction, sana14_multiplicity}.



In this paper, we explore the idea that the low-mass component was itself a bNS merger remnant, based on the following two motivations. (1) The mass of $2.6\msun$ is naturally produced from the coalescence of the known bNS systems, whose total masses are in the range from 2.5 to $2.9\msun$ \citep{LVC_GW170817_parameters, farrow19_bNS_mass_distribution} and possibly up to $\sim$3.4$\msun$ \citep{LVC20_bNS_GW190425}. (2) Only a small fraction of the bNS merger remnants need to participate in the 2nd-generation merger. The rate of GW190814-like mergers, estimated to be $\mc{R}_{\rm 190814}\simeq 7^{+16}_{-6}\rm\, Gpc^{-3}\, yr^{-1}$ \citep[90\% confidence interval,][]{abbott20_GW190814}, is much less than that of bNS mergers. From the two detected events GW170817 and GW190425\footnote{This event could also be from a NS-BH merger (see the discovery paper for a discussion). Since its remnant may also undergo a 2nd-generation merger, we include it in our rate analysis (although this corresponds to only a small difference in rate, see Fig. \ref{fig:rate_ratio}). Our calculations are weakly affected by the mass of the GW190425 remnant being slightly larger than $2.6\msun$. }, the bNS merger rate is estimated to be $\mc{R}_{\rm bns} \simeq 1090^{+1720}_{-800}\rm\,Gpc^{-3}\,yr^{-1}$ \citep[][different methods yield slightly different answers]{LVC20_bNS_GW190425}. Therefore, the GW190814-implied rate may be explained if only a fraction (in the range $0.06\%$ to $3\%$, see \S\ref{sec:rate_ratio}) of bNS merger remnants coalesce again with another BH.

There are a number of ways such 2nd-generation mergers can occur.


The first possibility is that a bNS merger occurred in a dense star cluster and then the remnant was dynamically captured by a more massive BH. Since direct GW-capture in single-single scattering is extremely inefficient due to the small cross section \citep{samsing20_single_single_encounters}, the more likely mechanism is an exchange during a binary-mediated encounter. However, in the case that the $23\msun$ BH is in a binary, its companion most likely has mass $\gg 2.6\msun$, and it is difficult for a $2.6\msun$ low-mass object to break apart the more massive binary and get captured. This scenario strongly favours equal-mass ratio components \citep{rodriguez19_dynamical_formation, ye20_NSBH_mergers, samsing20_mass_gap_BH_in_cluster}. Even when the low-mass BH is successfully captured by an exchange process, the resulting separation is likely too wide for the two BHs to merge within a Hubble time. As a rough representation of such asymmetric mass-ratio binaries, the NS-BH merger rate from globular clusters is estimated to be in the range 0.01 to 0.06$\,\rm Gpc^{-3}\,yr^{-1}$ \citep{ye20_NSBH_mergers} \citep[see also][]{arcasedda20_nuclear_star_cluster}, and the rate from dynamic assembly in young massive and open clusters is found to be of the same order \citep{fragione20_bhns_merger} \citep[but see][]{rastello20_young_massive_cluster}.

The second possibility is the ``double GW merger" scenario \citep{samsing19_doubleGW}, which postulates that a tight NS-NS binary scatters off a massive BH. Such binary-single scattering has been extensively studied in the past \citep{heggie75_binary_review, hut83_binary_single, hills91_binary_single, hills92_hard_binaries, sigurdsson93_binary_single, fregeau04_fewbody, samsing14_binary_single_GW}. During a resonant scattering where the whole system stays bound for much longer than the initial orbital period of the bNS system, the inner binary may be driven to merge rapidly and the subsequent remnant likely stays bound to the massive BH. However, this scenario requires that the initial bNS system has a merger time much shorter than the Hubble time \citep{samsing19_doubleGW}, and the probability that such a compact bNS system happens to have a close resonant encounter with a massive BH is likely too small to explain the inferred rate of GW190814.

A third possibility is that all three members were initially in a hierarchical triple system of three massive stars. The 23$\,M_\odot$ BH formed first with nearly zero spin \citep{fuller19}. The members of the inner binary each made a NS (the whole system survived the natal kicks), and common-envelope evolution or secular perturbation from the tertiary brought the two NSs sufficiently close so as to merge into a 2.6$\,M_\odot$ BH \citep{toonen16_triple, tauris17}. However, subsequently, it is unclear how to bring the low-mass remnant BH closer to the massive one. One possibility is that the binary BH (bBH) system is embedded in the accretion disk of an active galactic nucleus (AGN), the gas accretion onto the bBH system may operate to shrink the orbit \citep[e.g.,][]{stone17_bbh_AGN, bartos17_agn_assisted, mckernan18_bbh_AGN, yang19_hierarchical_mergers_ANG, yang20_bbh_in_AGN}, but a potential problem is that this scenario may lead to large BH spins aligned with the AGN disk. Secular perturbation by the supermassive BH's tidal field may subsequently excite large eccentricity in the remnant bBH system and lead to merger, but most likely at a very low rate \citep{fragione19_triple_near_SMBH}. Another possibility is to invoke an addition body in a hierarchical quadruple system in the field, with either ``2+2" or ``3+1" configuration such that after the bNS merger, the remaining three bodies undergo chaotic evolution to generate the 2nd-generation merger, but such fine-tuned situation may occur at a very low rate \citep{safarzadeh20_quadruple, fragione20_quadruple, liu20_quadrupole}.

In this work, we propose a new mechanism under the triple scenario where the bNS merger process imparts a natal kick on the remnant BH which leaves it in a low angular momentum orbit around the massive tertiary BH such that they can merge within a Hubble time. We show that up to about 1\% of the bNS mergers occurring in triple systems may give such 2nd-generation bBH merger, which is potentially consistent with the rate of GW190814-like sources.

This paper is organized as follows. In \S\ref{sec:rate_ratio}, we estimate the \textit{ratio} between the LIGO-inferred merger rate of GW190814-like systems and that of bNS in the local Universe. Detailed calculations of the formation probability of 2nd-generation mergers and comparison with observations are presented in \S\ref{sec:model}. The main uncertainty of our model --- whether current observations allow a large fraction of bNS mergers to occur in triples --- is discussed in \S\ref{sec:concern}. Our model provides a number of prediction which are presented in \S\ref{sec:predictions}. The main results are summarized in \S\ref{sec:summary}.




\section{Observed Rate Ratio}\label{sec:rate_ratio}

\begin{figure}
\centering
\includegraphics[width=0.47\textwidth]{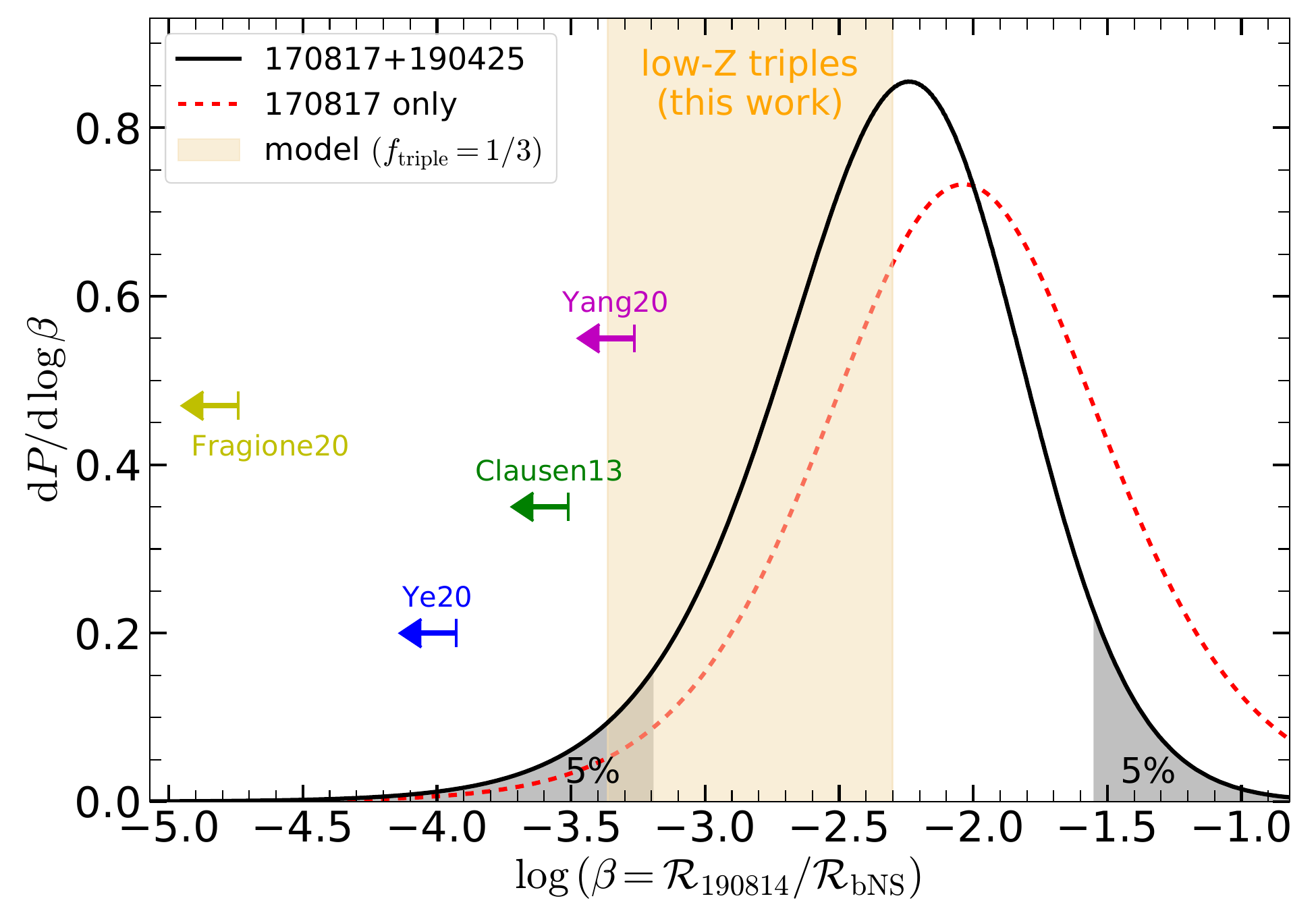}
\caption{The probability density function (PDF) for the rate ratio $\beta=\mc{R}_{190814}/\mc{R}_{\rm bns}$ as inferred from LIGO observations. The black solid line (our fiducial case) considers both GW170817 and GW190425 as bNS mergers, whereas the red dashed line only includes GW170817. The grey shaded regions are the lower and upper 5\% percentiles of the probability distribution. The vertical band (in light-orange color) shows the prediction from our low-metallicity triple scenario, under the assumption that $f_{\rm triple}=1/3$ of all bNS mergers occur in triple systems. The colored arrows show the 90\% upper limits based on the predicted rates of GW190814-like events from other possible scenarios in the literature. 
}
\label{fig:rate_ratio}
\end{figure}

Here, we estimate the probability density function (PDF) for the ratio $\beta=\mc{R}_{190814}/\mc{R}_{\rm bns}$ between the volumetric rate of GW190814-like events ($\mc{R}_{190814}$) and that of bNS mergers ($\mc{R}_{\rm bns}$), both measured in the local Universe. This PDF $\d P/\d \beta$ is given by the product distribution for the two independent random variables $\mc{R}_{190814}$ and $1/\mc{R}_{\rm bns}$, which we estimate below.

Since the errors of the LIGO-estimated rates are dominated by Poisson statistics \citep{abbott20_GW190814, LVC20_bNS_GW190425}, we approximate the PDF for the expectation number of detections $\mc{N}=\mc{R}VT$ (from the surveyed space-time volume $VT$) by $\d P/\d \mc{N}\propto \mc{N}^{k-1/2}\mr{e}^{-\mc{N}}/k!$, where $k=1$ for each of the relevant cases ($\mc{R}_{190814}$, $\mc{R}_{170817}$, and $\mc{R}_{190425}$), and the factor of $\mc{N}^{-1/2}$ is from Jeffrey's prior \citep{abbott20_GW190814}. From the median values of $\bar{\mc{R}}_{190814}=7\rm\, Gpc^{-3}\, yr^{-1}$ \citep{abbott20_GW190814}, $\bar{\mc{R}}_{\rm 170817}=760\rm\, Gpc^{-3}\, yr^{-1}$, and $\bar{\mc{R}}_{\rm 190425}=460\rm\, Gpc^{-3}\, yr^{-1}$ \citep{LVC20_bNS_GW190425}, we obtain the effective surveyed space-time volumes $VT=1.2/\bar{\mc{R}}$ for each of these three events (``1.2" is the median of $\d P/\d \mc{N}$). We consider both GW170817 and GW190425 as bNS mergers, because the component masses of GW190425 are not far from those of GW170817 and the nature of the merging objects makes little practical difference in our model. Thus, the PDF of the total bNS merger rate from the sum of the two is given by a convolution of the two individual PDFs
\begin{equation}
    {\d P\over \d \mc{R}_{\rm bns}} = \int_0^{\mc{R}_{\rm bns}} \d \mc{R}_1 {\d P\over \d \mc{R}_1} \left.{\d P\over \d \mc{R}_2}\right|_{\mc{R}_{\rm bns}-\mc{R}_1},
\end{equation}
where we have written $\mc{R}_{1} = \mc{R}_{170817}$, $\mc{R}_{2} = \mc{R}_{190425}$ for brevity. We then calculate the PDF for the inverse of the total bNS merger rate $\d P/\d \mc{R}_{\rm bns}^{-1}=\mc{R}_{\rm bns}^2\d P/\d\mc{R}_{\rm bns}$. Finally, the PDF of the rate ratio $\beta=\mc{R}_{190814}/\mc{R}_{\rm bns}$ is given by
\begin{equation}\label{eq:rate_ratio_pdf}
    {\d P\over \d \beta} = \int_0^\infty {\d \mc{R}_{3}\over \mc{R}_3} {\d P\over \d \mc{R}_{3}} \left.{\d P\over \d \mc{R}_{\rm bns}^{-1}}\right|_{\beta/\mc{R}_3},
\end{equation}
where we have written $\mc{R}_{3} = \mc{R}_{190814}$ for brevity. We find the 90\% confidence interval for the rate ratio to be in the range $0.064\%<\beta<2.8\%$.

The PDF as given by eq. (\ref{eq:rate_ratio_pdf}) is shown in a black solid line in Fig. \ref{fig:rate_ratio}. Alternatively, if we only consider GW170817 as bNS merger (simply replacing $\d P/\d \mc{R}_{\rm bns}$ by $\d P/\d \mc{R}_{170817}$), the resulting PDF for the rate ratio is shown in the red dashed line, which is rather similar to the black solid line.

Later in \S \ref{sec:model}, we assume that a fraction $f_{\rm triple}$ of the bNS mergers in the \textit{entire} Universe originated from low-metallicity triple systems that give rise to a merging bNS plus a massive tertiary. The precise threshold for low metallicity (perhaps $\sim$$0.1Z_\odot$) is rather uncertain but we impose it such that high-mass BHs $\gtrsim20\msun$ are efficiently generated. We then compute the probability $\lara{f_{\rm m}}$ of having a 2nd-generation merger by averaging over all possible triple configurations and kick velocities, which gives $0.1\% \lesssim \lara{f_{\rm m}}\lesssim 1.5\%$ (see Table \ref{tab:probability}). The scenario proposed here may be consistent with observations provide that $f_{\rm triple}\gtrsim 10\%$ (see \S \ref{sec:model}). We directly compare the model-predicted product $f_{\rm triple} \times \lara{f_{\rm m}}$, as shown in a vertical band in Fig. \ref{fig:rate_ratio} for $f_{\rm triple}=1/3$ as a representative value, with the observed rate ratio $\beta=\mc{R}_{190814}/\mc{R}_{\rm bns}$ in the local Universe.

We note that this comparison would only be accurate provided that the 2nd-generation bBH mergers from triples have similar delay-time distribution (DTD) as bNS mergers. However, we show later that the typical delay time for the 2nd-generation mergers is about a Hubble time (dominated by the bBH inspiral time after the bNS merger inside the triple, see eq. \ref{eq:delay_time_distribution}). Thus, the DTD of 2nd-generation bBH mergers is significantly shallower than the DTD of all bNS mergers, which is $\d P/\d t_{\rm d}\propto t^{-1}$ or steeper, as indicated by the declining deposition rate of radioactive elements in the solar neighborhood \citep[e.g.,][]{Hotokezaka15_Pu,beniamini20_rprocess_mixing} as well as the declining rate of short gamma-ray bursts at low redshift \citep{Wanderman15}. This means that the majority of bNS mergers occurred in the high-redshift Universe, and along with them, there are more 2nd-generation bBH systems generated that will merge only significantly later, at low redshift. The consequence is that the required value of $f_{\rm triple}$ to explain a given observed merger rate ratio in the local Universe, becomes smaller. In that sense, our choice for $f_{\rm triple}$ explained in later sections, is conservative.

We also show in Fig. \ref{fig:rate_ratio} the predictions from other possible scenarios in the literature.

For instance, \citet{fragione20_quadruple} studied the evolution of 2+2 quadruple systems where the binary that contains two NSs merge into a BH in the ``mass gap" first. The subsequent kick on the remnant BH triggers its interaction with the other binary system. If the remnant BH undergoes exchange with one of the members in the binary, then it may generate a merger that contains a mass-gap object. The rate of such mergers is found to be $10^{-2}\rm\, Gpc^{-3}\,yr^{-1}$ in the most optimistic case. We note that the 90\% upper limit as given by the $\d P/\d\mc{R}^{-1}_{\rm bns}$ distribution is $(\mc{R}^{-1}_{\rm bns})_{90}=1.8\times10^{-3} \rm\, Gpc^{3} \, yr$. Therefore, even taking the maximum rate of $\mc{R}\sim 10^{-2}\rm\, Gpc^{-3}\,yr^{-1}$ from \citet{fragione20_quadruple}, we obtain the 90\% upper limit for the rate ratio $\beta_{90} = \mc{R}\times (\mc{R}^{-1}_{\rm bns})_{90} = 1.8\times10^{-5}$ (as shown by the yellow arrow in Fig. \ref{fig:rate_ratio}), significantly below the most probable interval for $\beta$. \citet{clausen13_globular_cluster_BHNS} and \citet{ye20_NSBH_mergers} studied the dynamical formation of BH-NS mergers in globular clusters and found $\mc{R}_{\rm BHNS}\sim 0.17\rm\,Gpc^{-3}\,yr^{-1}$ and $\sim 0.06\rm\,Gpc^{-3}\,yr^{-1}$, respectively. We use their BH-NS rates as an upper limit for GW190814-like events (which have a more stringent requirement that the bNS merger remnants merge again with another BH), because the number density of bNS merger remnants is much smaller than that of NSs \citep{samsing20_mass_gap_BH_in_cluster}. The corresponding $\beta_{90}$ are shown in green and blue arrows.

The AGN-assisted merger scenario is more flexible because mass growth due to accretion may make $2.6\msun$ objects from normal NSs of initial mass $\approx 1.4\msun$. The 23$\msun$ component of GW190814 is also expected have gained mass through accretion, which would have to be stochastic in angular momentum orientation so as not to spin up the BH. In a recent study by \citet{yang20_bbh_in_AGN}, in their most optimistic case when allowing for up to 0.7 times the Bondi infall rate, they found a merger rate density of $3.1\rm\,Gpc^{-3}\,yr^{-1}$ for binaries which contain a member in the ``mass gap" (defined as 2.2--$5\msun$ by the authors). However, only about $10\%$ of these mergers have a mass ratio $q\sim 0.1$ and secondary component mass $m\sim 2.6\msun$ similar to GW190814. This means $\mc{R}\simeq 0.3\rm\, Gpc^{-3}\,yr^{-1}$ and $\beta_{90}\simeq 5\times10^{-4}$ as shown by a magenta arrow.

We conclude that the combination of the mass ratio and inferred rate of GW190814 is challenging to explain in previous models where a quantitative prediction of the formation rate has been presented. Our new scenario is described in the following section.

\section{The Model}\label{sec:model}
In this section, we first schematically describe our model in \S\ref{sec:sketch}, followed by detailed calculations of the occurrence rate of 2nd-generation mergers in \S\ref{sec:fmerger} and \S\ref{sec:SMA_distribution}. The final results are presented in \S\ref{sec:results}.

\subsection{Schematic Picture}\label{sec:sketch}
\begin{figure}
\centering
\includegraphics[width=0.47\textwidth]{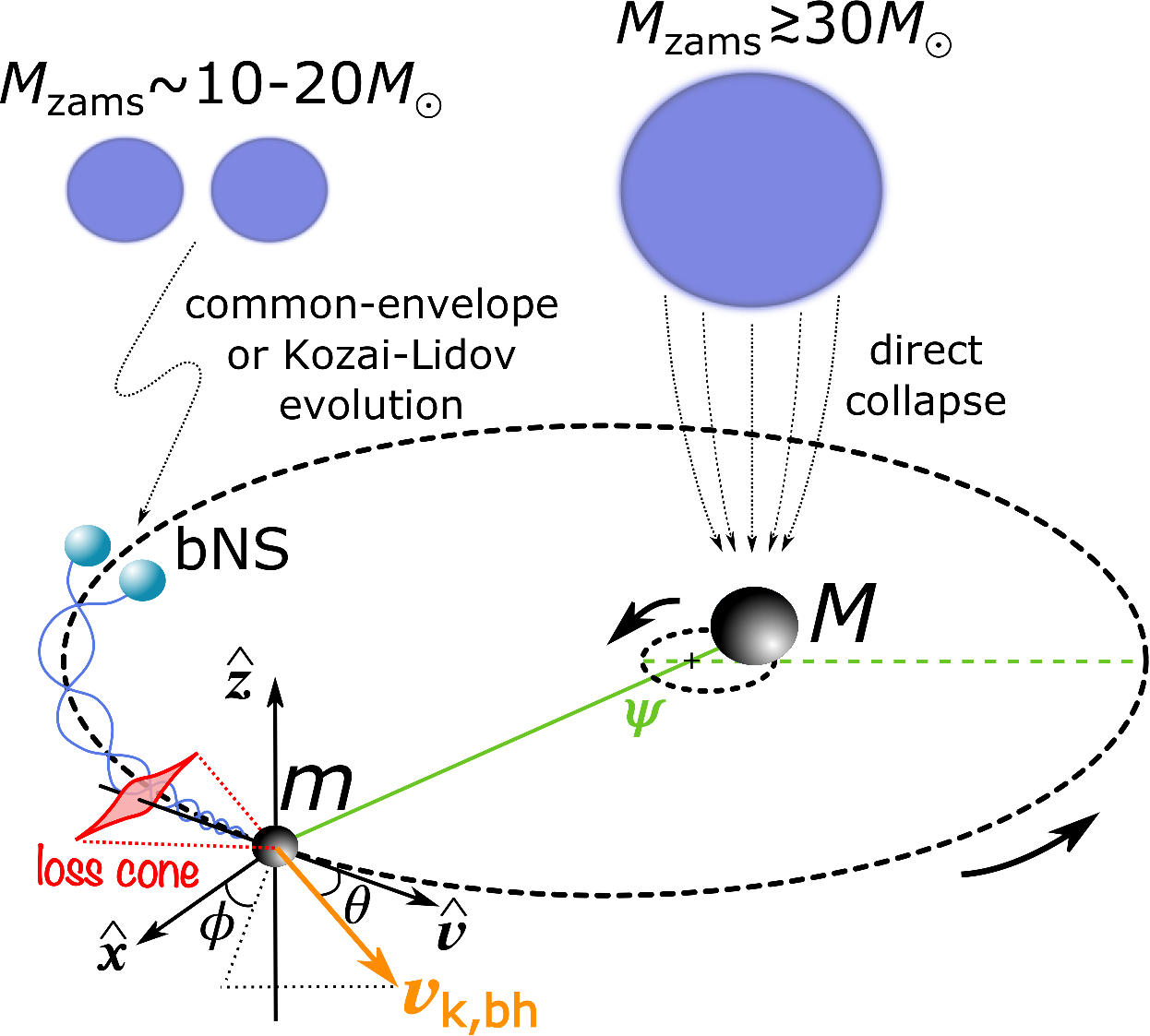}
\caption{Sketch of the model for GW190814 (not to scale). The initial hierarchical triple system consists of two stars of $10$--$20\msun$ in the inner orbit and a tertiary star of mass $\gtrsim 30\msun$, all on the main-sequence. The most massive tertiary evolves into a BH by direct collapse of either the entire star or the helium core. The inner system evolves into a compact NS-NS binary via common-envelope evolution or Kozai-Lidov (KL) excitation under the perturbation of the tertiary.
At outer orbital phase $\psi$ (the true anomaly), the two NSs first merge into a $2.6\msun$ remnant BH. A combination of baryonic ejecta and GW emission during the merger gives the remnant BH a kick $\b{v}_{\rm k,bh}$ in the $(\theta,\ \phi)$ direction, where $\theta$ is the angle between $\b{v}_{\rm k,bh}$ and the pre-merger orbital speed $\b{v}$ of the bNS system and $\phi$ is the azimuthal angle from the $\hx$-$\hv$ plane. The unit vector $\hx$ is within the pre-kick outer orbital plane, perpendicular to $\hv$, and pointing away from the massive BH. The unit vector $\hz$ is along the direction of the angular momentum of the outer orbit. We find that a fraction (up to $1\%$) of the cases give rise to 2nd-generation bBH merger within a Hubble time, and this happens when the kick direction is in the ``GW loss cone" (shown in red) so as to cancel the pre-kick orbital angular momentum.
}
\label{fig:sketch}
\end{figure}

In this subsection, we describe the main features and assumptions of the model, which is schematically shown in Fig. \ref{fig:sketch}.

The initial triple system consists of three massive main-sequence stars formed at low metallicity (perhaps $\lesssim 0.1Z_\odot$). A low metallicity is motivated by the large mass of the primary in GW190814. It is naturally realized assuming that the systems under consideration were formed many Gyrs ago when the Universe was less metal enriched. As we will show below, this is consistent with the typical time between formation and final 2nd-generation merger. The inner binary consists of two stars in the mass range $\sim$10 to $20\msun$, each with radius $\sim$5$\,R_\odot$ at zero age main-sequence (ZAMS). The semi-major axis (SMA) of the inner orbit is larger than about $20R_{\odot}$ so as to avoid strong Case A mass transfer which may lead to rapid merger before any supernova explosion \citep{sana12_binary_interaction, schneider15_close_binary_interaction}; the stellar radius expands by a factor of about 2 to 3 during main-sequence evolution \citep{klencki20_stellar_radius_low_Z}. We consider the most likely situation in which the mass of the tertiary is comparable to the total mass of the inner binary \citep{moe17_binary_distribution}. To make a stable hierarchical system, the SMA of the outer orbit, denoted as $a_0$, must be larger than that of the inner binary by a factor of $\sim3$ or more \citep[e.g.,][]{kiseleva96_stability, mardling01, silsbee17_triple_stability}. The above requirements only allow sufficiently wide outer orbits with $a_0\gtrsim 0.3\rm\, AU$.

The star in the outer orbit, the most massive one ($\gtrsim 30\msun$), evolves into a BH by direct collapse of either the entire star or the helium core. The mapping between ZAMS mass to the final BH mass is highly uncertain \citep[e.g.,][]{fryer12_BH_masses, spera15_BH_masses, woosley19} \citep[see also][for a discussion in light of the LIGO bBH population]{LVC19_bbh_population}, depending on the details of wind mass loss and convective mixing \citep[e.g.,][]{klencki20_stellar_radius_low_Z}, interaction with the companion (the inner binary for our case) in the H-shell or core-He burning phase, and supernova physics. Thus, we consider a number of possible BH masses of $10$, $20$, and $30\msun$, which are all possible under the low-metallicity condition considered here. Efficient rotational coupling removes angular momentum from the core to the envelope and then to the wind \citep{fuller19}, so the BH spin is likely small. Accretion feedback may unbind the outer envelope and further reduce the BH spin \citep[e.g.,][]{batta19_accretion_feedback}. The BH birth may be associated with a natal kick (of e.g., a few tens $\rm km\, s^{-1}$), which may disrupt the triple system if the outer orbit is very wide (with separation $\gtrsim100\rm\,AU$). We focus on the systems that survive this kick.

Later on, the inner binary, due to the close separation of the two stars, goes through the common-envelope channel of producing two NSs that will merge in less than a Hubble time \citep[see Fig. 1 of][]{tauris17}. It is also possible that the inner bNS system is driven to a high eccentricity by the secular perturbation from the tertiary \citep[e.g.,][]{antognini14_triple_eccentricity}. Along with each supernova explosion, there is a natal kick on the newly made NS, and we focus on the cases where the two NS kicks didn't unbind the inner binary (the unbound fraction due to these kicks is effectively taken into account by our normalization based on bNS merger rate, see \S\ref{sec:rate_ratio}).

In the following, we mainly consider the effects of the second-born NS's kick on the outer orbit. This is because as explained next, the effect of the kick due to the first-born NS on the outer orbit is comparable to that of the second. Furthermore, these NS formation kicks do not significantly affect our final rate of 2nd-generation mergers (see Table \ref{tab:probability}), because in the cases that lead to a bBH merger, the outer orbit is mainly affected by the much larger kick on the remnant BH during the bNS merger.

Kinematic studies of the Galactic bNS systems show that the second-born NS, on average, receives a kick of $\lesssim 50\rm\, km/s$ \citep[e.g.,][]{Shaviv2005,schwab10_low_kick,beniamini16_bNS_kicks}, which is significantly smaller than that for young isolated pulsars \citep{hobbs05}. The favored explanation for such a small kick is that the pre-explosion star was heavily stripped by the close-orbit NS companion \citep{tauris15_ultra_stripped_SN}, and such an ultra-stripped supernova has a very small ejecta mass of order $0.1\msun$ and an explosion energy of order $10^{50}\rm\, erg$ \citep{suwa15_ultra_stripped}. During the explosion, the center of mass of the inner binary system experiences a sudden change in velocity\footnote{The first NS likely had received a much larger kick of the order $\sim300\rm\, km/s$ \citep[e.g.,][]{hobbs05}, which is typical of the kicks leading to the formation of isolated pulsars. However, since this is shared with a more massive ($\gtrsim10\msun$) main-sequence companion star, the change in the binary's center-of-mass velocity is reduced by a factor of 10. Overall, the first NS kick has a similar effect on the outer orbit as the second NS kick. The same is likely to apply to the natal kick received by the tertiary BH. More detailed discussion is provided in Appendix C.} of $\b{v}_{\rm k,bns} \approx \b{v}_{\rm k,ns}/2$ as a result of linear momentum gain of $\mns \b{v}_{\rm k,ns}$, which is due to the kick on the newly born NS as well as the mass lost from the binary (the contributions from these two processes are likely comparable, see \citealt{beniamini16_bNS_kicks}).
We assume the direction of $\b{v}_{\rm k,bns}$ to be isotropically distributed and take the amplitude distribution to be log-normal with mean value $\mr{log}\,\bar{v}_{\rm k, ns}$ and standard deviation $\sigma_{\mr{log}\,v_{\rm k, ns}} = 0.3$ dex in log space (corresponding to a factor of 2 larger or smaller). As compared to the Maxwellian, a log-normal distribution better describes the large scatter of kick speeds as inferred from the Galactic bNS systems \citep[cf. Fig. 16 of][]{tauris17}. We consider three cases of $\bar{v}_{\rm k,bns}=10$, 20, and $30\rm\, km/s$. For simplicity, we assume the pre-kick outer orbit to be circular\footnote{This assumption is conservative because, for a given outer SMA $a_0$, non-zero eccentricity $e_0>0$ leads to a larger probability for the 2nd-generation mergers (cf. Fig. \ref{fig:fmerger2D}).}, and then the kick modifies the SMA and eccentricity of the outer orbit in a predictable way \citep{hills83, kalogera96}.

After the GW merger time of the inner binary, the two NSs coalesce into a low-mass BH (or an extremely massive NS, as our model is independent of the nature of the merger remnant). The baryonic ejecta, GWs, and possibly neutrinos lost during the bNS merger carry linear momentum, and correspondingly the remnant BH receives a kick $\b{v}_{\rm k,bh}$. We aim to calculate the probability that the low-mass BH is kicked into a sufficiently low angular momentum orbit that it merges with the tertiary (the massive BH) within a Hubble time.

If we ignore tidal disruption, in the limit of nearly equal masses, the component of the kick due to GW emission is given by $v_{\rm k, bh}\approx 287 (1-q)\rm\, km/s$ \citep{gonzalez07_spin_kick}, where $q \leq 1$ is the mass ratio of the bNS system. The majority of the Galactic bNS systems have $0.9\lesssim q<1$, except for the PSR J1913+1102 system which has $q\simeq 0.8$ \citep{ferdman20_low_mass_ratio} but this source likely only represents a small fraction of the volumetric bNS merger rate \citep{andrews20_mass_ratio}. We see that the kick due to GW is typically small, $\lesssim 30\rm\, km/s$ for the majority of the cases. On the other hand, the kick due to baryonic ejecta depends on the mass ratio, NS equation of state, and possibly hydrodynamic effects of the material outside the BH. \citet{dietrich17} carried out numerical relativity simulations of bNS mergers for different mass ratios and equations of state, and the linear momentum carried away by the dynamical ejecta (their Table 3) gives kick amplitude in the range $25$--$130\rm\, km/s$ for $0.8\lesssim q\lesssim 1$, often subjected to simulation resolution (since only $10^{-3}$--$10^{-2}\msun$ is dynamically ejected).

A much larger amount ($10^{-2}$--$10^{-1}\msun$) of material is bound to the remnant BH. The bound material experiences numerous shocks and quickly forms a nearly circular accretion disk. Numerical simulations show that a fraction of order unity of the disk material will be ejected as a result of viscous evolution and helium recombination at later time \citep[e.g.,][]{siegel17_merger_disk, fernandez19_merger_disk}. However, the linear momentum carried away by the disk wind is highly uncertain, and a fractional asymmetry of $mv_{\rm k,bh} v_{\rm w}/E_{\rm w} \simeq 16\%$ is required to generate $v_{\rm k,bh}=100\,\rm km/s$ for typical wind velocity $v_{\rm w} = 0.1c$, kinetic energy $E_{\rm w}=10^{51}\rm\, erg$, and BH mass $m=2.6\msun$. It is also unclear whether the neutrino emission from the short-lived proto-NS carries significant linear momentum (depending on the neutrino transport in the magnetized NS interior), and a small fractional asymmetry of 0.5\% can generate $v_{\rm k,bh}=100\,\rm km/s$ if neutrinos carries away $E_\nu = 3\times10^{53}\rm\, erg$ of energy.

Given the uncertainties and potentially large case-to-case variations, we consider log-normal amplitude distribution with three different mean values at $\bar{v}_{\rm k, bh}=50,\ 100,\ 150\,\rm km/s$ in log space, and fix the standard deviation at $\sigma_{\mr{log}\,v_{\rm k,bh}} = 0.3$ dex. We found that Maxwellian distributions with similar median amplitudes give qualitatively similar results. We assume the direction of $\b{v}_{\rm k,bh}$ to be isotropically distributed, because the orbital orientation of the inner binary may be easily changed by the torque from the outer orbit (the total angular momentum is dominated by the outer orbit).

In the next subsection, we calculate in detail the fraction of triple systems that give rise to 2nd-generation mergers.

\subsection{Probability for 2nd-generation mergers}\label{sec:fmerger}

Our calculation follows two steps: {\bf (1)} For an initial outer SMA $a_0$ (assuming a circular orbit), we give a random kick $\b{v}_{\rm k,bns}$ (due to the formation of the second NS in the binary, as described above) to the center of mass of the bNS system with a log-normal amplitude ($v_{\rm k, bns}$) distribution and an isotropic direction ($\mu_0\equiv \cos\theta_0, \phi_0$) distribution, and then obtain the post-kick outer SMA $a$ and pericenter $\rp$; {\bf(2)} For each outer orbit as specified by $a$ and $\rp$, we consider that the bNS merger occurs at a random orbital phase $\psi$ and that the remnant BH receives a kick $\b{v}_{\rm k, bh}$ with a log-normal amplitude ($v_{\rm k, bh}$) distribution and an isotropic direction ($\mu \equiv \cos\theta$, $\phi$) distribution. Thus we obtain the post-merger SMA $a'$ and $\rp'$. If the GW merger time of the final bBH system is less than $10\rm\, Gyr$, then we record a successful 2nd-generation merger. The mass loss along with each of the kicks is only of order 1\% of the total mass and is hence ignored. The masses of the two BHs are denoted as $M$ (the more massive one) and $m$, and they stay fixed throughout the orbital evolution. We consider three different cases of $M=10,\ 20,\ 30\msun$ and fix $m=2.6\msun$ for all cases. In the following, we calculate in detail the response of an orbit with general eccentricity and semi-major axis to a kick.
This is applicable to Step 2 of our calculation. Step 1 is then a simpler special case of this calculation, where the orbit is initially circular.

Consider a binary of masses $M$ and $m$ in an orbit with SMA $a$ and eccentricity $e=1-\rp/a$. The angle $\psi$ denotes the true anomaly, which is the angle between the current position of the orbiting object and the location of pericenter. At orbital phase $\psi$, the binary separation is given by
\begin{equation}
    r = {a(1-e^2)\over 1 + e\cos\psi},
\end{equation}
and the relative velocity between the two objects is
\begin{equation}\label{eq:pre-kick_speed}
    v = \sqrt{G(M+m) (2/r - 1/a)},
\end{equation}
where $G$ is Newton's constant. The specific orbital energy $\eps$ and specific angular momentum $\ell$, defined as the corresponding total values divided by the reduced mass $Mm/(M+m)$, are given by
\begin{equation}
    \eps = -G(M+m)/2a,\ \ \ell^2 = G(M+m)a(1-e^2),
\end{equation}
The angular momentum is along the $\hat{\b{z}}$ direction. The angle between the (relative) velocity vector $\b{v}$ and the (relative) position vector $\b{r}$ is given by
\begin{equation}
    \sin\alpha = {\ell\over vr} = {1+e\cos\psi\over \sqrt{1 + 2e\cos\psi + e^2}},
\end{equation}
and $\alpha < \pi/2$ when the two objects are moving away from each other, and $\alpha > \pi/2$ otherwise.

When the system is at orbital phase $\psi$, the object $m$ suddenly experiences a randomly oriented general kick $\b{v}_{\rm k}$ whose amplitude is drawn from a log-normal PDF $\d P/\d \vk$. The direction of $\b{v}_{\rm k}$ is specified by the angle $\theta$ to the pre-kick orbital velocity $\b{v}$ and an azimuthal angle $\phi$ wrt. the $\hat{\b{x}}$-$\hat{\b{v}}$ plane (i.e. $\phi=0$ when $\b{v}_{\rm k}$ is in this plane), where the base vector $\hat{\b{x}}$ is within the pre-kick orbital plane, perpendicular to $\b{v}$, and pointing outwards away from the center of mass. The geometry is illustrated in Fig. \ref{fig:sketch}. Thus, in the $\hx$-$\hv$-$\hz$ coordinate system, one can write the position vector $\b{r} = r(\sin\alpha\, \hx + \cos\alpha\, \hv)$ and kick velocity $\bvk = \vk(\sin\theta \cos\phi\, \hx + \cos\theta\, \hv + \sin\theta \sin\phi\, \hz)$. Under the assumption of no mass loss, the updated orbital energy is given by
\begin{equation}
    \eps' = {1\over 2}(\b{v} + \bvk)^2 - {G(M+m)\over r} = \eps + {1\over 2}\vk^2 + v\vk\cos\theta.
\end{equation}
We focus on the the bound cases with $\eps'<0$, where the updated SMA $a'$ is
\begin{equation}\label{eq:aRatio}
    {a\over a'} = 1 - (y^2 + 2y\cos\theta) (2a/r - 1).
\end{equation}
The change in specific angular momentum is $\b{r}\times\bvk$, so the updated eccentricity is given by
\begin{equation}\label{eq:AngMmtRatio}
    {a'(1-e'^2)\over a(1-e^2)} = {y^2\sin^2\theta \sin^2\phi\over \sin^2\alpha} + [1 + y(\cos\theta - \mr{cot}\,\alpha \sin\theta \cos\phi)]^2,
\end{equation}
where we have denoted $y\equiv \vk/v$. The above can be summarized by the following mapping\footnote{As for Step 1, since the pre-kick orbit is assumed to be circular, the mapping can be described by
$$(a_0,\ e_0=0)\xrightarrow[\mbox{NS kick}]{(y_0, \theta_0, \phi_0)} (a,\ e),$$
where we take pre-kick separation $r = a_0$ (without dependence on orbital phase), $\alpha=\pi/2$, kick amplitude $y_0 \equiv v_{\rm k,bns}/v_0$, and pre-kick orbital speed $v_0 = \sqrt{G(M+m)/a_0}$.}
\begin{equation}
    (a,\ e)\xrightarrow[\mbox{BH kick}]{(\psi, y, \theta, \phi)} (a',\ e'),
\end{equation}
which means that a kick as defined by the parameters $(\psi,\ y,\ \theta,\ \phi)$ maps the initial orbit described by $(a,\ e)$ into a different orbit $(a',\ e')$. Based on symmetry, we only need to consider $\psi\in (0,\ \pi)$, $\theta\in (0,\ \pi)$, and $\phi\in (0,\ \pi)$.

We are interested in the rare cases where the post-kick orbit is highly eccentric $1-e'\ll 1$ and the merger time for such a system is given by \citep{peters64}
\begin{equation}
\begin{split}
    t_{\rm GW} &\approx {3c^5\over 85G^3 Mm(M+m)} a'^{1/2} [a'(1-e'^2)]^{7/2} \\
    &= 1.15\mr{\,Gyr} {\msun^3\over Mm(M+m)} {a'^{1/2} [a'(1-e'^2)]^{7/2}\over (10^{11}\mr{\,cm})^4}.
\end{split}
\end{equation}
We require $t_{\rm GW} < 10\rm\, Gyr$ such that that the post-kick angular momentum (or pericenter distance) is much smaller than the pre-kick value. This means the RHS of eq. (\ref{eq:AngMmtRatio}) is much less than unity. Making use of eqs. (\ref{eq:aRatio}) and (\ref{eq:AngMmtRatio}), we obtain the requirement for 2nd-generation merger
\begin{equation}\label{eq:y_quadr}
\begin{split}
   & {y^2\sin^2\theta \sin^2\phi\over \sin^2\alpha} + [1 + y(\cos\theta - \mr{cot}\,\alpha \sin\theta \cos\phi)]^2 < f_{\rm c}, \\
    & f_{\rm c}\equiv  4.6\times10^{-2} \left(Mm(M+m)\over 1.17\times10^3\msun^3\right)^{2\over 7}
    {\left(a/a'\right)^{1\over 7} \over a_{\rm AU}^{8/7} (1-e^2)},
\end{split}
\end{equation}
where the mass product term has been normalized by $M=20\msun$ and $m=2.6\msun$, and $a_{\rm AU}\equiv a/\mr{AU}$. For simplicity, we ignore the factor $(a/a')^{1/7}\simeq 1$ (see eq. \ref{eq:aRatio}) in the following, because it only affects the results very weakly. Then, the RHS of the above inequality is set by the initial orbital parameters $f_{\rm c}(a, e)$ (for fixed masses), and the LHS depends on the orbital phase ($\psi$) at which the kick occurs as well as the kick amplitude ($y\equiv \vk/v$) and direction ($\theta$ and $\phi$). Instead of using a brute-force Monte Carlo method to find the merger fraction for each set of $(a,\ e)$, it is better to analytically narrow down the range of kick direction and amplitude possible for mergers.

\begin{figure}
\centering
\includegraphics[width=0.47\textwidth]{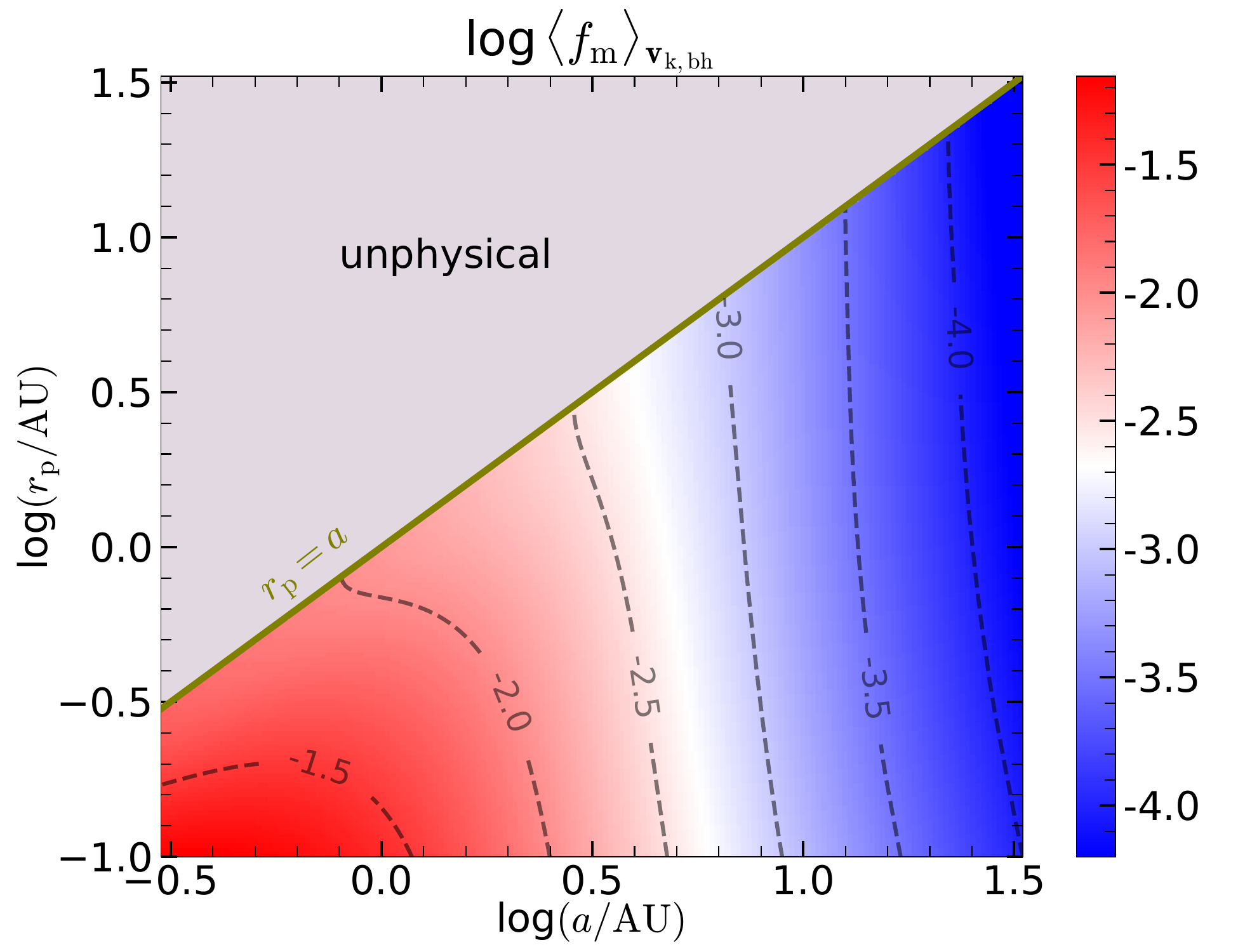}
\caption{The fraction of bNS merger remnants $\lara{f_{\rm m}}_{\b{v}_{\rm k,bh}}$ that will coalesce with the tertiary BH as a function of the SMA $a$ and pericenter distance $r_{\rm p}$ of the outer orbit right before the bNS merger. We have averaged over the distributions of the amplitude (log-normal) and direction (isotropic) of the kick $\b{v}_{\rm k,bh}$ received by the low-mass remnant BH after the bNS merger. The case shown here is for $M=20$, $m=2.6\msun$, $\bar{v}_{\rm k,bh}=100\rm\, km/s$. 
}
\label{fig:fmerger2D}
\end{figure}

Since the LHS of the inequality (\ref{eq:y_quadr}) is a quadratic function of $y$, it is easy to show that, for each $f_{\rm c}<1$ (valid for the entire parameter space considered in this work), solutions of $y$ only exist when $\cot\theta < \cot \alpha \cos\phi - \sqrt{f_{\rm c}^{-1}-1}\sin\phi/\sin\alpha$. We only need to consider $\mu\equiv \cos\theta$ in the following range
\begin{equation}
    -1<\mu < {\xi\over \sqrt{1 + \xi^2}},\ \ \xi\equiv \cot \alpha \cos\phi - \sqrt{f_{\rm c}^{-1}-1}{\sin\phi\over \sin\alpha},
\end{equation}
and the 2nd-generation merger solution lies in between
\begin{equation}
\begin{split}
    y_\pm = {1 \pm \sqrt{(\eta+1)f_{\rm c} - \eta} \over \gamma(\eta+1)},
\end{split}
\end{equation}
where
\begin{equation}
    \eta  \equiv \left(\sin\theta \sin\phi \over \gamma \sin\alpha\right)^2,\ \  \gamma\equiv \cot\alpha \sin\theta \cos\phi - \cos\theta.
\end{equation}
We also require the post-kick orbit to be bound, and from eq. (\ref{eq:aRatio}) and $a'>0$, we obtain
\begin{equation}
    y < y_{\rm unb} = \sqrt{\mu^2 + (2a/r-1)^{-1}} - \mu.
\end{equation}
Therefore, the final range of $y$ that leads to post-kick GW merger is
\begin{equation}\label{eq:y_range}
    \max(0,\ y_-) \equiv y_{\rm min} < y <  y_{\rm max}\equiv \min(y_+,\ y_{\rm unb}).
\end{equation}
Then, for a given pre-kick orbit $(a,\ \rp)$, the probability of 2nd-generation merger for a particular set of $(\psi,\ \theta,\ \phi)$ is given by an integral of the (log-normal) PDF of the kick amplitude $\int \d v_{\rm k,bh} (\d P/\d v_{\rm k,bh})$ in the range given by $y_{\rm min}v<v_{\rm k,bh} < y_{\rm max} v$, where $v$ is given by eq. (\ref{eq:pre-kick_speed}). Then, it is straightforward to average the probability over the PDFs of the other parameters,
\begin{equation}\label{eq:prior}
    {\d P\over \d \psi} = {1\over \pi} {(1-e^2)^{3/2}\over (1 + e\cos\psi)^2},\ {\d P \over \d\mu} = {1\over 2},\ {\d P \over \d\phi} = {1\over \pi},
\end{equation}
over the ranges of $\psi\in (0,\ \pi)$, $\mu\in (-1,\ 1)$, and $\phi\in (0,\ \pi)$.

The kick-averaged probability of 2nd-generation merger $\lara{f_{\rm m}}_{\b{v}_{\rm k,bh}}$ as a function of the pre-kick orbital parameters ($a,\ \rp$) is shown in Fig. \ref{fig:fmerger2D}. The 2nd-generation merger fraction can reach $1\%$ or higher for $a\lesssim$ a few AU and generally decreases for larger pre-kick outer SMA. Note that we have ignored the cases where the post-kick orbit is unbound $\eps'>0$ but the pericenter $\rp'$ is sufficiently close to allow for GW capture in one pericenter passage and then merger. As we describe next, this situation is highly improbable. For nearly parabolic orbits, the GW energy loss in one pericenter passage is given by \citep{peters63}
\begin{equation}
    \delta E_{\rm GW} = {85\pi G^{7/2}\over 12\sqrt{2} c^5} {(M m)^2\sqrt{M+m} \over \rp'^{7/2}} = {1.4\times10^{46}\mr{\,erg}\over r_{\rm p,9}'^{7/2}},
\end{equation}
where we have taken $M=20\msun$ and $m=2.6\msun$. When the post-kick orbit is unbound, the typical excessive orbital energy is of the order $m v_{\rm k,bh}^2/2\sim 6\times10^{46}\, (v_{\rm k,bh}/50\rm\, km\,s^{-1})^2\rm\, erg$. We see that the post-kick angular momentum has to be extremely close to zero, with $\rp'\lesssim 10^{-4}\rm\, AU$, in order for GW capture to happen. We find the contribution from GW capture to 2nd-generation mergers to be less likely than GW inspiral in bound orbits by a factor of $10^{-2}$.

\begin{figure}
\centering
\includegraphics[width=0.47\textwidth]{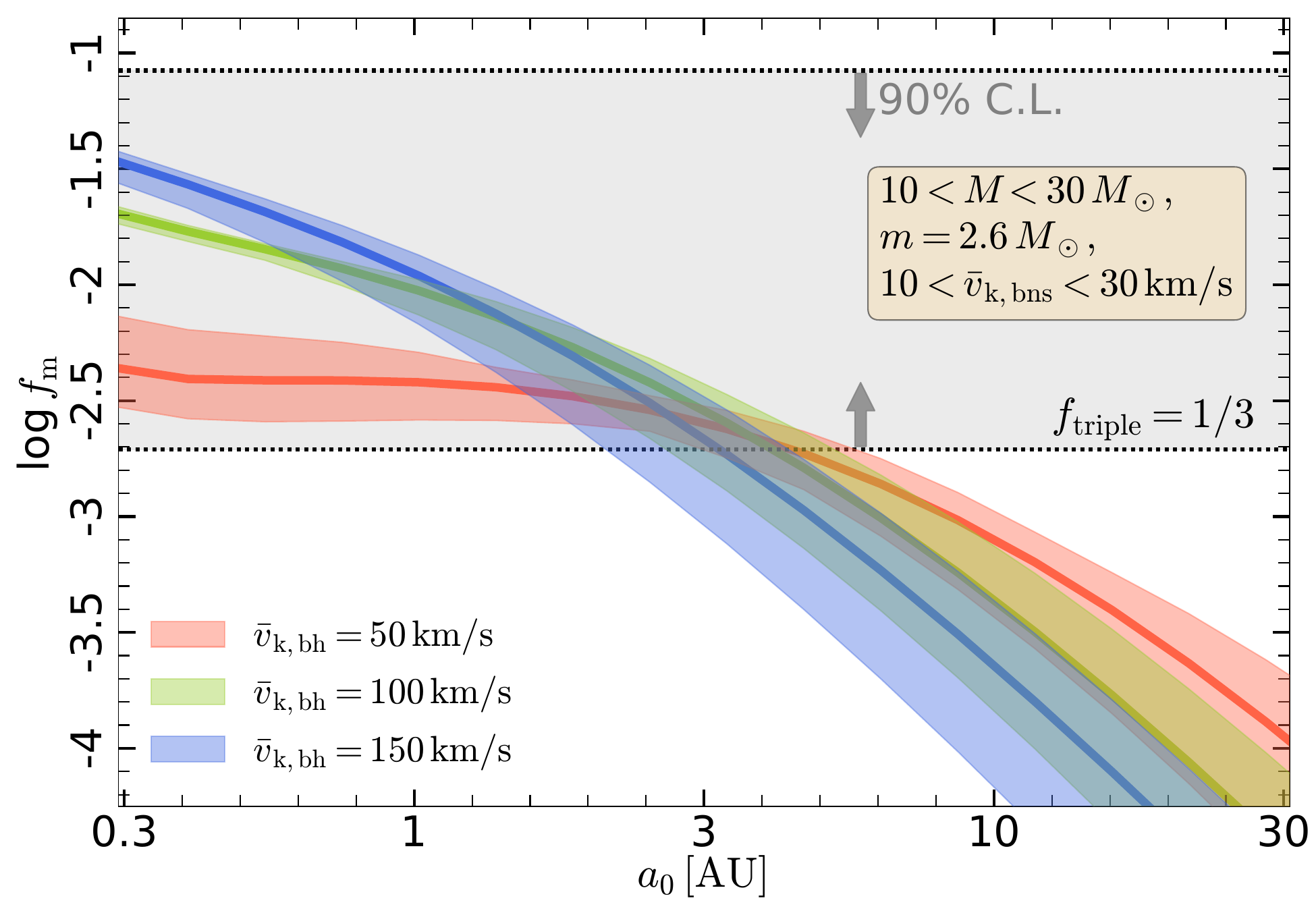}
\caption{The fraction of the bNS mergers that generate a 2nd-generation merger, $f_{\rm m}$, as a function of initial SMA of the outer orbit $a_0$ right before the birth of the second NS. We have averaged over the distributions of the two kicks on the bNS system and on the low-mass remnant BH. The three colored bands are for different median BH kick speeds $\bar{v}_{\rm k,bh}=50,\ 100,\ 150\rm\, km/s$. Each band is for tertiary BH mass in the range $10<M<30\msun$ and the median kick amplitude on the bNS system in the range $10<\bar{v}_{\rm k,bns}<30\,\rm km/s$. The solid lines in the middle of each band is for $M=20\msun$ and $\bar{v}_{\rm k,bns}=20\rm\, km/s$. The dependence on $\bar{v}_{\rm k,bns}$ is rather weak for $a_0\lesssim 3\rm\,AU$, since the outer orbit is only weakly affected by this kick. The black dotted lines indicate the 90\% confidence interval for the observationally allowed volumetric rate ratio between GW190814-like systems and bNS mergers, provided that $f_{\rm triple}=1/3$ of the bNS merger rate is contributed by triples.
}
\label{fig:fmerger1D}
\end{figure}

Taking into account Step 1 is now straightforward. For an initially circular orbit, and given the PDF of the kick $\b{v}_{\rm k,bns}$ on the bNS system due to birth of the second NS, we calculate the semi-major axis, and eccentricity at the beginning of Step 2. The double-averaged (over both $\b{v}_{\rm k,bns}$ and $\b{v}_{\rm k,bh}$) merger fraction, denoted as $f_{\rm m} (a_0)$, is only a function of the initial SMA $a_0$ of the outer orbit. This is shown in Fig. \ref{fig:fmerger1D}, for a number of choices of the tertiary BH mass $10<M<30\msun$, median kick speed on the bNS system $10<\bar{v}_{\rm k,bns}<30\rm\, km/s$, and the median kick speed on the remnant BH $50<\bar{v}_{\rm k,bh}<150\rm\, km/s$. For instance, for the fiducial case of $M=20\msun$, $\bar{v}_{\rm k,bns}=20\rm\, km/s$, and $\bar{v}_{\rm k,bh}=100\rm\, km/s$, a fraction $f_{\rm m}\simeq 0.3\%$ (1\%) of the bNS mergers will give rise to a 2nd-generation merger, if the outer SMA is at $a_0=3\rm\,AU$ ($1\rm\, AU$).

Our model assumes that the formation rate of triples including a merging bNS plus a massive tertiary represents a fraction $f_{\rm triple}$ of all bNS mergers.
Earlier in \S\ref{sec:rate_ratio}, we have calculated the PDF for the ratio $\beta \equiv \mc{R}_{190814}/\mc{R}_{\rm bns}$ between the LIGO-inferred merger rate of GW190814-like events and that of bNS mergers in the local Universe. Therefore, we also show in Fig. \ref{fig:fmerger1D} the 90\% confidence interval for $\beta/f_{\rm triple} = \mc{R}_{190814}/(f_{\rm triple} \mc{R}_{\rm bns})$, which is directly compared to $f_{\rm m}$, the fraction of triple systems at a given outer SMA $a_0$ that give 2nd-generation mergers. We find that, for $f_{\rm triple}=1/3$ (see \S\ref{sec:discussion_predictions} for a discussion of this choice), our model-predicted 2nd-generation merger rate from triples is potentially in agreement of the rate of GW190814-like events, provided that a large fraction of triples have outer SMA $a_0\lesssim 3\rm\, AU$. In the next section, we discuss the distribution of the outer SMA.

\subsection{Distribution of the outer SMA}\label{sec:SMA_distribution}

The final piece of information needed in our model is the physical distribution of the outer SMA $\d P/\d a_0$ right before the birth of the second NS, and this distribution strongly depends on the initial SMA distribution of triples in the main-sequence stage as well as possible orbital migration due to stellar interactions, e.g., Roche-lobe overflow, common envelope, tidal dissipation, 3-body secular dynamics. A thorough discussion of the interplay among these poorly understood processes across a wide range of metallicities is beyond the scope of this work \citep[see e.g.,][]{postnov14_binary_evolution, toonen16_triple}. This potential shortcoming of the current model requires careful studies in future works. In the following, we proceed with the simplest, power-law prescription for $\d P/\d a_0$.

We are only interested in triple systems in which bNS mergers occur. We emphasize that we are normalizing the fraction of 2nd-generation mergers by the rate of bNS merging in triples, so those triple systems that do not end up contributing to the latter are ignored for our purpose. We assume their outer SMAs $a_0$ before the formation of the second NS to have the following power-law distribution,
\begin{equation}\label{eq:SMA_distribution}
    {\d P\over \d \mr{log}\,a_0}\propto a_0^{-p}, \ \ a_{\rm 0,min}< a_0 < a_{\rm 0,max},
\end{equation}
where $p=0$ corresponds to {\:O}pik's law (flat in log space) and we choose an upper limit of $a_{\rm 0,max}=100\rm\, AU$ such that the natal kicks associated with the birth of the two NSs do not disrupt the inner and outer orbit at high probability. The exact value of the upper limit do not affect our results, since the probability of generating a 2nd-generation merger is negligibly small at $a_0\gtrsim 100\rm\, AU$.

Various studies of massive star multiplicity indicate $p\gtrsim 0$ during the main-sequence stage \citep[e.g.,][]{sana12_binary_interaction, kobulnicky14_period_distribution, almeida17_period_distribution}. However, it is unclear how this is modified by subsequent stellar interaction at close separations. Our results will be mainly affected by outwards orbital migration that may lead to a larger outer SMA $a_0$ than at zero-age main-sequence, because the fraction of 2nd-generation merger is generally smaller for larger $a_0$ (cf. Fig. \ref{fig:fmerger1D}). We note that at low metallicities with reduced wind mass loss such that the entire triple system can retain a large fraction of its initial mass, the outer SMA can at most increase by a factor of order unity \citep{postnov14_binary_evolution} \citep[see also Fig. 4 of][]{rodriguez18_triple}, as confirmed by our preliminary simulations using $\mathtt{MESA}$-$\mathtt{binary}$ \citep{Paxton2019}.

Without considering orbital migration, a rough estimate of the lower limit of $a_{\rm 0,min}$ can be obtained by the following argument. The inner binary we are considering have initial masses between $\sim$10 and $20\msun$ (since NS end products are required). These stars expand to about $10$--$20R_\odot$ at the end of their main-sequence, and lower metallicity stars expand less due to reduced opacity \citep{klencki20_stellar_radius_low_Z}. To avoid mergers due to rapid Case-A mass transfer during main-sequence stage \citep{sana12_binary_interaction, schneider15_close_binary_interaction}, we require the inner SMA to be more than $20R_\odot$. This constrains the SMA of the outer orbit, because for the triple system to be hierarchically stable, the ratio between the outer and inner SMAs $Y=a_{\rm out}/a_{\rm in}$ must be greater than a critical value \citep{kiseleva96_stability, silsbee17_triple_stability}
\begin{equation}\label{eq:triple_stability}
    Y_{\rm c} \simeq {3.7\over Q_{\rm out}} - {2.2 \over Q_{\rm out} + 1} + {1.4\over Q_{\rm in}} {Q_{\rm out} - 1 \over Q_{\rm out} + 1},
\end{equation}
where $Q_{\rm in} = (m_1/m_2)^{1/3}$, $Q_{\rm out}=(m_{12}/m_3)^{1/3}$, $m_1$ and $m_2 (>m_1)$ are the component masses of the inner binary, $m_{12}=m_1 + m_2$ is the total mass of the inner binary (at main-sequence stage), $m_3$ is the mass of the tertiary, and we have assumed the inner and outer orbits to be close to circular (appropriate for such compact systems). For typical cases with $Q_{\rm in}\simeq Q_{\rm out} \simeq 1$, we obtain $Y_{\rm c}\simeq 3$, and therefore secular stability requires the outer SMA to be more than $3\times 20R_\odot\simeq 0.3\rm\, AU$. Due to large uncertainties in the stellar and dynamical evolution of the triple system, we consider two cases of $a_{\rm 0,min}= 0.3\rm\, AU$ and $1\rm\,AU$.

The inner bNS system can be driven to extremely high eccentricity by the perturbation of the massive BH tertiary by the Kozai-Lidov (KL) mechanism \citep{shappee13_eccentric_kozai, antognini14_triple_eccentricity, toonen16_triple, silsbee17_triple_stability, liu18_triple_merger, rodriguez18_triple, fragione19_bhns_merger_triples, hamers19_triple_rate, liu19_merger_eccentricity}, provided that the inclination between the inner and outer orbits is close to $90^{\rm o}$. Roughly speaking, KL excitation operates when the ratio $Y$ between the SMAs of the outer and inner orbits is in the range from 5 to about 100 \citep[larger SMA ratios are allowed if the outer orbit is highly eccentric,][]{rodriguez18_triple, liu19_merger_eccentricity}. The upper limit is because when $Y\gtrsim 100$, the secular perturbation of the tertiary is too weak to affect the inner orbit. The lower limit is because when $Y\lesssim 5$ the triple system becomes unstable (eq. \ref{eq:triple_stability}, but for $Q_{\rm out}\simeq 0.5$ due to smaller mass ratio $m/M\sim 0.1$). The KL excitation only operates in a narrow inclination window close to 90$^{\rm o}$, and the window is narrower for larger inner SMA (and hence larger outer SMA), because the inner orbit must have a higher eccentricity for GW merger to happen within a Hubble time. It can be shown under the secular quadrupole approximation that the KL merger fraction $f_{\rm KL}$, defined as the fraction of the inclination space (from 0 to 180$^{\rm o}$) taken by the KL merger window, roughly scales as $f_{\rm KL}\propto a_{\rm out}^{-2/3}$ \citep{liu18_triple_merger}, which corresponds to $p=2/3$. We conclude that the bNS mergers from the KL channel strongly prefer smaller outer SMAs.

A caveat for the KL mechanism is that the inner binary may be driven to merge during the main-sequence stage, if the initial orbital inclination is close to orthogonal. It is also possible that the inner binary is driven to high eccentricity by the KL mechanism but without a main-sequence merger, and subsequently the inner orbit tidally circularizes and undergoes common-envelope evolution to generate a bNS merger. The above consideration motivates us to consider a range of $0\leq p\leq 2/3$, where $p=0$ corresponds to the limit where the inner system undergoes effectively isolated common-envelope evolution whereas $p=2/3$ represents the KL channel with a strong preference for smaller outer SMAs.


\renewcommand{\arraystretch}{1.2}
\begin{table*}
\centering
\caption{The fraction ($\%$) of bNS systems that undergo a 2nd-generation merger, averaging over the power-law distribution of the outer SMA $a_0$ (eq. \ref{eq:SMA_distribution}) before the birth of the second NS. We fix the mass of the two BHs to be $M=20\msun$ and $m=2.6\msun$. In our fiducial case, the median kick amplitude on the bNS system at the birth of the second NS is $\bar{v}_{\rm k,bns}=20\rm\, km/s$. The values in parentheses are for the results from $\bar{v}_{\rm k,bns}=10\rm\, km/s$  (for weak kicks), which show that the NS formation kicks do not significantly affect our  final results. For different choices of outer SMA distributions (minimum SMA $a_{\rm 0,min}$ and power-law index $p$) and median kick amplitudes $\bar{v}_{\rm k,bh}$ for the remnant BH, we find the fraction of 2nd-generation mergers to be in the range from $0.1\%$ to $1\%$. This is potentially in agreement with the observed volumetric rate ratio between GW190814-like systems and bNS mergers, provided that a significant fraction (more than $\sim 10\%$) of the bNS mergers in the Universe originate from low-metallicity triples.
}
\label{tab:probability}
\begin{tabular}{|c|c|c|c|c|} 
\hline
\multirow{2}{*}{$a_{\rm 0,min}$}     & \multirow{2}{*}{$p$}         &   \multicolumn{3}{c}{fraction of 2nd-generation merger $\lara{f_{\rm m}}_{a_0}$ ($\%$)} \\ 
\multicolumn{2}{c}{} &  $\bar{v}_{\rm k,bh}=50\rm\,km/s$    &   $\bar{v}_{\rm k,bh}=100\rm\,km/s$   & $\bar{v}_{\rm k,bh}=150\rm\,km/s$ \\
\hline
\multirow{3}{*}{0.3 AU} & 0 & 0.19 (0.16)    &   0.45 (0.43)   & 0.58 (0.58)\\

\multicolumn{1}{c}{} & 0.3 & 0.26 (0.23)    &   0.72 (0.69)   & 0.98 (0.97)\\

\multicolumn{1}{c}{} & 2/3 & 0.33 (0.28)    &   1.0 (0.99)   & 1.5 (1.4)\\
\hline
\multirow{3}{*}{1 AU} & 0 & 0.13 (0.12)    &   0.19 (0.18)   & 0.17 (0.17)\\

\multicolumn{1}{c}{} & 0.3 & 0.18 (0.16)    &   0.28 (0.28)   & 0.26 (0.27)\\

\multicolumn{1}{c}{} & 2/3 & 0.24 (0.21)    &   0.39 (0.39)   & 0.38 (0.39)\\
\hline
\end{tabular}
\end{table*}
\renewcommand{\arraystretch}{1}

\subsection{Results}\label{sec:results}
The final merger fraction after averaging over the $a_0$ distribution
\begin{equation}
    \lara{f_{\rm m}}_{a_0}\equiv \int_{a_{\rm 0,min}}^{a_{\rm 0,max}} \d a_0 {\d P\over \d a_0} f_{\rm m}(a_0),
\end{equation}
is shown in Table \ref{tab:probability} for a number of choices of $a_{\rm 0,min}$, $p$, and $\bar{v}_{\rm k,bh}$. We find that the $a_0$-averaged fraction is in the range of $\lara{f_{\rm m}}_{a_0}\in (0.1\%,\ 1\%)$, and this is potentially in agreement with the observed rate ratio between GW190814-like systems and bNS mergers as measured by LIGO \citep{LVC20_bNS_GW190425, abbott20_GW190814}, provided that a significant fraction (more than $\sim 10\%$) of the bNS mergers in the Universe originate from low-metallicity triples.

In the Appendix, we show the Markov-Chain Monte Carlo (MCMC) sampling of the posterior for the 8 parameters involved in our model: initial outer SMA $a_0$, kick on the bNS system $(v_{\rm k, bns}, \mu_0\equiv\cos\theta_0, \phi_0)$, orbital phase $\psi$ when bNS merger occurs, and kick on the remnant BH $(v_{\rm k, bh}, \mu\equiv\cos\theta, \phi)$. The case shown in Fig. \ref{fig:posterior} is for flat prior in $\mr{log}\,a_0$ or $p=0$ in eq. (\ref{eq:SMA_distribution}). We see that most 2nd-generation bBH mergers occur in the cases of very tight outer orbit $a_0\lesssim$ a few AU and when the kick on the remnant BH is either in the opposite direction of the orbital velocity near $\mu\simeq -1$ or close to the orbital plane near $\phi\simeq 0$ (or $\pi$) so as to cancel the pre-kick orbital angular momentum. The ``GW loss cone" is shaped like a flying saucer, as shown in Fig. \ref{fig:sketch}. For the vast majority of our cases that will undergo a 2nd-generation merger, the delay time wrt. the bNS merger time is near 10 Gyr and only a very small fraction have short delay time $t_{\rm d}\ll 10\rm\, Gyr$. These short-delay-time cases correspond to very small post-BH-kick angular momentum or pericenter distance $\rp'$.

Analogous to the ``loss cone" picture in tidal disruption events \citep[e.g.,][]{stone20_TDE_rate}, the probability of reaching below certain pericenter distance $\rp'$ is given by $\d P/\d\, \mr{log}\,\rp'\propto \rp'$. Since the GW merger time, or the delay time, scales as $t_{\rm d}\propto \rp'^{7/2}$, we find the delay time distribution to be
\begin{equation}\label{eq:delay_time_distribution}
    {\d P\over \d\,\mr{log}\,t_{\rm d}}\propto t_{\rm d}^{2/7},
\end{equation}
which means that there are a lot more long-delay-time cases than the ``standard" delay time distribution that is flat in $\mr{log}\,t_{\rm d}$. This is consistent with our hypothesis that GW190814-like systems come from old, low-metallicity stellar population. On the short-delay end of the distribution, we find the fraction with delay time $\lesssim 100\,\rm yr$ to be roughly $10^{-16/7}\simeq 5\times10^{-3}$, which means that it is possible but highly unlikely to observe ``double mergers" where bNS and then bBH mergers are detected in quick succession \citep[see][for preliminary searches]{veske20_double_GW_merger}.

When the frequency of GWs from the bBH system reaches $f_{\rm GW}=10\rm\, Hz$ (the edge of LIGO band), the orbit is almost circular with eccentricity $e_{\rm m}\ll 1$ and the SMA at this time is given by \citep{wen03_merger_eccentricity}
\begin{equation}
    a_{\rm m} \simeq \left[{G(M+m)\over (\pi f_{\rm GW})^2}\right]^{1/3} \simeq 9.7\times10^{-6}f_{\rm GW,10Hz}^{-2/3}\rm\, AU,
\end{equation}
for $M=20\msun$ and $m=2.6\msun$. The ``merger eccentricity" when the system enters the LIGO band can be estimated by $e_{\rm m} \simeq (a_{\rm m}/ 1.76\rp')^{19/12}$ in the limit $e_{\rm m}\ll 1$ \citep{peters64}. Thus, we obtain the merger eccentricity distribution to be
\begin{equation}
    {\d P\over \d\,\mr{log}\,e_{\rm m}}\propto e_{\rm m}^{-12/19},
\end{equation}
for $e_{\rm m}\ll 1$.
We find that the majority of the bBH merger cases in our model have eccentricities $e_{\rm m}\simeq 3\times10^{-6}f_{\rm GW,10Hz}^{-19/18}$, and only a small fraction of the order $10^{-3}$ of them will have $e_{\rm m}\gtrsim 0.1$ in the LIGO band. However, future GW detectors such as LISA \citep{amaroseoane17_lisa} and TianQin \citep{luo16_tianqin, mei20_tianqin} will likely detect GW190814-like sources at much lower frequencies a long time $t_{\rm GW}$ before the merger, where
\begin{equation}\label{eq:merger_time}
    t_{\rm GW}\simeq 2.2\times10^2\mr{\,yr}\, (\mc{M}/6\msun)^{-5/3} f_{\rm GW,10mHz}^{-8/3},
\end{equation}
$\mc{M} = (Mm)^{3/5}/(M+m)^{1/5}\simeq 6\msun$ is the chirp mass of GW190814 \citep{abbott20_GW190814} and $f_{\rm GW,10mHz} = f_{\rm GW}/10\rm\, mHz$ is the GW frequency. At much lower frequencies, we estimate that 90\% (or 10\%) of GW190814-like events have eccentricity of $e\gtrsim 2\times10^{-3} f_{\rm GW, 10mHz}^{-19/18}$ (or $e\gtrsim 0.1 f_{\rm GW, 10mHz}^{-19/18}$). This is quite promising because LISA may be sensitive to eccentricities as low as $10^{-3}$ \citep{nishizawa16_LISA_eccentricity}.




\section{Discussion and Predictions}\label{sec:discussion_predictions}

In this section, we first discuss the biggest concern of our model: do current observations allow a large fraction of bNS mergers to come from low-metallicity triples? After addressing this potential concern, we provide a number of predictions that will be useful to test our model in the future.

\subsection{Fraction of bNS mergers in low-metallicity triples}\label{sec:concern}

We have shown that, in hierarchical bNS-BH triple systems, the fraction of the bNS merger remnants that undergo a 2nd-generation merger with the tertiary BH within a Hubble time is in the range from $0.1\%$ to $1.5\%$. For comparison, the observed volumetric rate ratio between GW190814-like systems and bNS mergers is in the range $0.06\%$ to $3\%$ (90\% confidence interval) as inferred from LIGO observations. Therefore, our model can potentially explain the rate of GW190814-like systems provided that a significant fraction (more than $\sim 10\%$) of bNS mergers in the Universe occur in triples that are born at low metallicity.

It is perhaps counter-intuitive, but such a large fraction of bNS mergers in triples is in fact allowed by current observations, despite the fact that none of the known bNS systems in our Milky Way have a massive tertiary.

The known Galactic bNS systems in the field (i.e., not in globular clusters) are produced by high-metallicity stellar population from which the generation of very massive BHs $\gtrsim 20\msun$ is highly suppressed. We know that more than half of the massive main-sequence binaries in the Milky Way have a tertiary \citep{sana14_multiplicity}. Thus, many of the Galactic bNS systems should have had a tertiary at the beginning of their evolution, and the outer orbits must have been dissociated in the past. This dissociation is more likely at higher metallicity because strong wind mass loss of the tertiary tend to widen the outer orbit and produce a lower mass compact remnant, and hence it is easier for the outer orbit to be disrupted by natal kicks.

The Galactic bNS systems are relatively young, with ages of less than 100 Myr \citep{beniamini19_DTD}. This immediately implies that a population of systems contributing to bNS mergers, formed early on in the evolution of our Galaxy, is not expected to be present in the known (radio-selected) Galactic bNS population. The observed delay-time distribution of the Galactic bNS population indicates that a large fraction of Galactic bNS merge within less than a Gyr \citep{beniamini19_DTD}. The majority of Galactic bNS were born much further in the past, when the star formation rate in the Galaxy was higher \citep{majewski93}. The combination of short merger time and declining formation rate suggests a declining rate of bNS mergers in the Galaxy \citep{Beniamini16_kicks_UFDs}. Indeed, this is also consistent with the declining deposition rate of r-process elements\footnote{We should note that bNS mergers are not the only possible source of r-process elements. Other rare core-collapse events, such as the formation of rapidly rotating strongly magnetized neutron stars \citep{mosta18_rproess_mhd_sn} or ``collapsars" responsible for long-duration gamma-ray bursts \citep{siegel19_collapsar_rprocess}, may also be responsible for r-process enrichment at low metallicity in the early Universe. } in the solar neighborhood as inferred from abundance ratios in radioactive isotopes \citep{Wallner15,Hotokezaka15_Pu}, as well as with the declining rate of short gamma-ray bursts at low redshift \citep{Wanderman15}. A declining rate of bNS mergers is consistent with the possibility of having additional formation channels of systems leading to bNS mergers at low metallicity, such as the one considered in this work. Indeed, if the overall rate of bNS forming at those early times was larger, the required fraction of systems involving a massive tertiary would be lower than the value of $f_{\rm triple}$ implied by directly comparing the merger rates $\mc{R}_{190814}$ and $\mc{R}_{\rm bns}$ in the local Universe.


Observations of r-process materials in a fraction of ultra-faint dwarf (UFD) galaxies \citep{ji16_rprocess_udfs, ji16_rprocess_abundances, roederer16_rprocess_udfs} indicate that the enrichment events happened at a rate of about once every $10^3$ core-collapse supernovae and within the first Gyr after the star formation \citep{beniamini16_rprocess_production}. If we assume bNS mergers dominate the production of r-process elements \citep[as indicated by many independent lines of evidence, e.g.,][]{hotokezaka18_rprocess_review}, then a potential problem is that the kick imparted on the center of mass of the bNS systems from the birth of either NSs may unbind a large fraction of them from the UFDs, which have typical escape speed of about $15\rm\, km\, s^{-1}$. { \cite{Beniamini16_kicks_UFDs} have shown that for the distribution of center of mass kicks as inferred from the Galactic bNS population (consisting of a number of ultra-striped supernovae with small kicks and little mass ejection), a fraction $\gtrsim 0.5$ of bNS systems are expected to remain confined and merge even in an UFD. Nonetheless, it is interesting to note that the scenario proposed in this work, will make confinement even easier. The existence of a massive tertiary, implies that the kick on the center of mass of the entire triple system is reduced by a factor of $\sim 10$.
This could lead to high r-process enrichment even in the most metal poor stars in UFD galaxies --- a signal that could be searched for once more complete abundance data from UFDs becomes available.}

\subsection{Predictions of our model}\label{sec:predictions}

\begin{figure}
\centering
\includegraphics[width=0.47\textwidth]{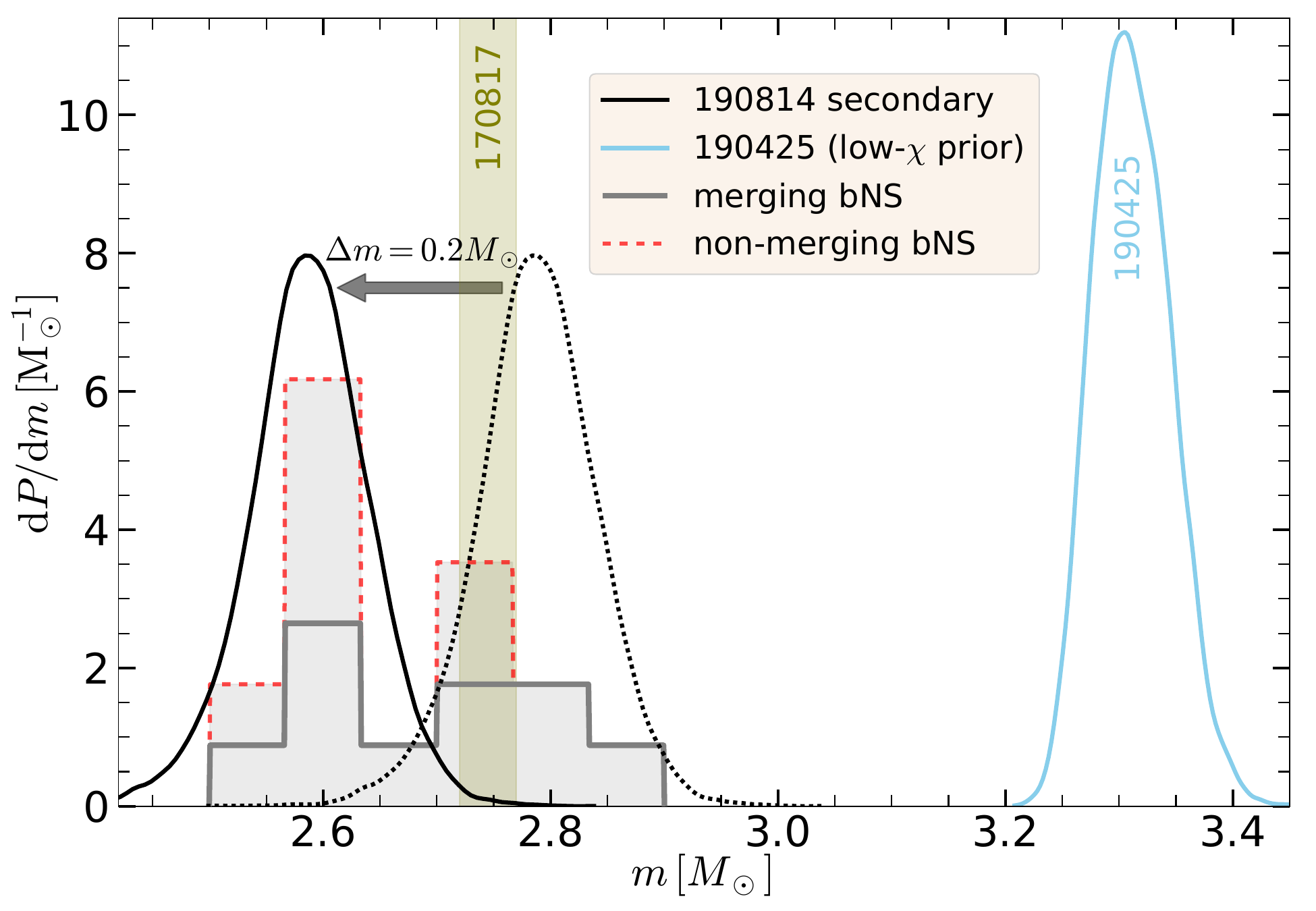}
\caption{Comparison between the secondary mass of GW190814 (black-solid line) and the total masses of known bNS systems, including the ones in the Milky Way (grey histogram), GW170817 (green-shaded region, $2.73^{+0.04}_{-0.01}\msun$ at 90\% confidence interval), and GW190425 (cyan-solid line). The Galactic bNS systems are divided into two groups: the ``merging" ones (grey-solid line) with GW inspiral time less than a Hubble time and the ``non-merging" ones (red-dashed line) with longer inspiral time. The posteriors for the total mass of GW170817 \citep{LVC_GW170817_parameters} or GW190425 \citep{LVC20_bNS_GW190425} are based on low-spin prior of $\chi<0.05$, because rapidly spinning NSs should have spun down during GW inspiral and
also the Galactic bNS population all have low spins \citep{zhu18_bns_low_spin_prior}. Under the hypothesis that the secondary of GW190814 was a bNS merger remnant, the total mass of the pre-merger bNS system was slightly larger than shown in the black-solid line, because GWs, neutrinos, and baryonic ejecta lost during the bNS merger reduced the total mass by $0.1\msun \lesssim \Delta m\lesssim0.3\msun$. We also show in a black-dotted line the distribution of the inferred total mass of the pre-merger bNS system by assuming a mass loss of $\Delta m=0.2\msun$ in the merger process.
}
\label{fig:bns_total_mass}
\end{figure}

Our model provides a number of testable predictions.

\begin{enumerate}
    \item[(1)] The spin of the secondary in all GW190814-like systems should be 0.6--0.7 \citep{gonzalez07_spin_kick, dietrich17}, because the majority of the angular momentum of the bNS system at merger is retained by the remnant BH. The posterior from modeling the LIGO waveform \citep{abbott20_GW190814} only constrains the spin of the secondary of GW190814 to be $0.53\pm0.3$ at 68\% confidence (see Fig. \ref{fig:secondary_spin}), but a more accurate spin measurement may be possible for higher signal-noise-ratio events in the future. According to our model, the secondary spin may be randomly oriented if the inner binary was driven to merge by the Kozai-Lidov mechanism, or preferentially aligned with the orbital angular momentum if the inner orbit was brought into alignment with the outer orbit by mass transfer when the tertiary fills its Roche lobe (which could happen when the tertiary was in the core He burning stage).
    \item[(2)] The component mass distribution of all LIGO detected sources should have a narrow peak between $2.5$ and $\sim$3.5$\msun$ as given by the total masses of the Galactic bNS systems and known bNS mergers (GW170817 and GW190425), provided that stellar evolution do not generate such BHs in the ``mass gap." In Fig. \ref{fig:bns_total_mass}, we show that the secondary mass of GW190814 is indeed similar to the total masses of known bNS systems, which supports the hypothesis that the secondary of GW190814 was a bNS merger remnant. We note that, even if the core-collapse pathway can generate mass-gap BHs, the triple scenario proposed in this work can still contribute a subset of GW190814-like events with unique secondary-mass distribution.
    \item[(3)] We estimate that 90\% (or 10\%) of GW190814-like events have eccentricity of $e\gtrsim 2\times10^{-3} f_{\rm GW, 10mHz}^{-19/18}$ (or $\gtrsim 0.1 f_{\rm GW, 10mHz}^{-19/18}$), which may be detectable by LISA/TianQin.
    \item[(4)] A significant fraction $(\gtrsim10\%)$ of bNS mergers should have signatures of a massive tertiary at a distance of about 1 AU in the GW waveform. The GW signal from the merging bNS system becomes observable when the frequency crosses the lower edge of the detector's frequency range $f_{\rm min}=f_{\rm min,Hz}\rm \, Hz$ at $t(f_{\rm min})\simeq 5.4\mr{\,d}\, f_{\rm min,Hz}^{-8/3}$ before the merger, where we have taken a fiducial chirp mass of $\mc{M}=1.2\msun$ in eq. (\ref{eq:merger_time}). Gravity from the tertiary causes an acceleration of the line-of-sight velocity of the bNS center of mass $\dot{v}_{\rm los}\sim GM/a^2$, where $M$ is the mass of the tertiary ($\approx$ the total mass) and $a$ is the SMA of the outer orbit. The acceleration over the merger time $t$ causes a non-linear cumulative Doppler phase variation in the GW waveform \citep{yunes11_GW_tertiary, meiron17_tertiary_GW_perturbation}
    \begin{equation}
        \Delta \Phi \sim {\dot{v}_{\rm los}t^2f_{\rm min}\over c} \sim 10^2\mr{\,rad} {M/20\msun\over a_\mr{AU}^{2}} f_{\rm min, Hz}^{-13/3}.
    \end{equation}
    We see that perturbation on the GW waveform from a $M=20\msun$ tertiary at a distance of $a=1\rm \,AU$ will be negligible for LIGO, since $\Delta \Phi\ll 1\rm \,rad$ for $f_{\rm min}=10\rm\,Hz$. However, detection of such a tertiary will be possible for future low-frequency ($f\lesssim1\rm\, Hz$) space-based observatories such as  LISA \citep{amaroseoane17_lisa}, DECIGO \citep{sato17_decigo}, TianGO \citep{kuns19_tiango}, and TianQin \citep{mei20_tianqin}.
    \item[(5)] Since the delay-time distribution for the 2nd-generation merger is dominated by the cases with $t_{\rm d}\sim 10\rm\, Gyr$ (eq. \ref{eq:delay_time_distribution}), the volumetric rate inferred from GW190814 constrains the number of bNS systems with a massive BH tertiary in our Galaxy: $\mc{R}_{190814}\times 10\mr{\,Gyr}/n_{\rm MW}\gtrsim 10^3$, where $\mc{R}_{190814}>1\rm\, Gpc^{-3}\,yr^{-1}$ \citep[at $95\%$ C.L.,][]{abbott20_GW190814} and $n_{\rm MW}\sim 10^7\rm\, Gpc^{-3}$ is the number density of Milky Way-like galaxies. In our model, roughly $1\%$ (or less) of the bNS-BH triples produce 2nd-generation mergers, so we conclude that the true number of bNS systems with a BH tertiary\footnote{Based on the delay-time distribution of the observed population of Galactic bNS systems, \citet{beniamini19_DTD} inferred the number of bNS systems (without a tertiary) in the Milky Way to be only slightly larger, on the order of $3\times 10^5$.} in the Milky Way is on the order of $10^5$. These NSs are no longer active in radio emission due to their old age, since they were formed in low metallicity environment in the distant past. One can also estimate that about $10^5\times1\mr{\,Myr}/10\mr{\,Gyr}\sim 10$ bNS systems will have merger time of $1\rm\, Myr$ or less (or GW frequency higher than about $1\rm\, mHz$). Such systems will be detectable by LISA/TianQin \citep{lau20_bNS_LISA_detections}, and the existence of a possible massive tertiary will be easy to unveil. It is also possible to detect the bNS-BH or the subsequent bBH systems by microlensing \citep[see e.g.,][]{wyrzykowski20_microlensing}.
\end{enumerate}

\section{Summary}\label{sec:summary}

We have proposed a model for the formation of the mysterious event GW190814 with component masses of $23\msun$ and $2.6\msun$, based on the idea of a 2nd-generation merger. Motivated by the fact that the mass of the secondary is very similar to the typical total mass of known bNS systems, we hypothesize that the secondary (most likely a low-mass BH) was made from a bNS merger, which was initially in a hierarchical triple with a massive tertiary BH. If the outer orbit is sufficiently tight at the time of bNS merger, then the bNS-merger remnant has a non-negligible probability to coalesce with the massive BH as a result of the natal kick imparted on the remnant in the bNS merger process. We calculated this probability in detail and found that about 0.1\% to 1\% of bNS mergers occurring in triples may give rise to a 2nd-generation merger. This is potentially in agreement with the observed ratio $0.06\% < \beta\equiv \mc{R}_{190814}/\mc{R}_{\rm bns} < 3\%$ (90\% confidence interval) between the rates of GW190814-like events and bNS mergers, provided that a significant fraction, $\gtrsim 10\%$, of bNS mergers occur in triple systems with outer SMA less than a few AU.

We suggest that these systems are from triples of massive stars formed at low metallicity (perhaps $\lesssim 0.1Z_\odot$) many Gyrs ago when the Universe was less metal-enriched. This is because massive stars near solar metallicity typically do not make BHs as heavy as the primary in GW190814, as demonstrated by empirical BH mass distribution in X-ray binaries \citep[e.g.,][]{ozel10_BH_mass_gap} as well as stellar evolution models \citep[e.g.,][]{belczynski16_lowZ}. Due to reduced wind mass loss during stellar evolution, low-metallicity triples retain a large fraction of the original total mass and hence experience less orbital expansion than high-metallicity systems. Thus, their outer orbits stay tight, which facilitates 2nd-generation mergers, because the probability of having 2nd-generation mergers generally decreases with the outer SMA. High-metallicity triples typically undergo significant orbital expansion due to strong mass loss and then may be easily disrupted by the natal kicks at the formation of the compact remnants \citep{rodriguez18_triple}. 

We argue that a significant fraction ($\gtrsim10\%$) of bNS mergers having occurred in triples is in fact allowed by current observations, despite the fact that none of the Galactic bNS systems has a massive tertiary. This is mainly based on two considerations. First, nearly all massive main-sequence stars are observed to be in multiple systems, and the average number of companions per system is about 2 \citep{sana14_multiplicity}. This is likely true at low metallicity as well, because the multiplicity fraction typically increases towards lower metallicity \citep{moe19_multiplicity}. Second, multiple lines of evidence (based on e.g., r-process abundance ratios and the redshift evolution of short gamma ray bursts) suggest that the rate of both bNS formation and merger has significantly declined over the evolution of our Galaxy. This leaves room for additional formation pathways for bNS mergers at low metallicity, such as the triple channel considered in this work. Even a relatively small fraction of bNS mergers occurring in triples in the distant past can produce a sufficiently large ratio between the rate of 2nd-generation bBH mergers today and the current rate of bNS mergers.

Our model provides many predictions and hence can be tested/falsified in the near future by looking for the following signatures:
\begin{itemize}
    \item[(1)] The spin of the secondary in  GW190814-like systems is 0.6--0.7.
    \item[(2)] The mass distribution of the secondary component of a large LIGO-source sample should have a narrow peak between $2.5$ and $\sim$3.5$\msun$ as given by the total mass of the known bNS systems, i.e., about half of the ``mass gap", from $\sim$3.5 to $\sim$5$\msun$, will stay empty (provided that stellar evolution do not generate such BHs).
    \item[(3)] About 90\% (or 10\%) of GW190814-like events will have eccentricities of $e\gtrsim 2\times10^{-3}$ (or $\gtrsim0.1$) near GW frequency of $10\, $mHz, which may be detectable by LISA/TianQin.
    \item[(4)] A significant fraction ($\gtrsim10\%$) of bNS mergers should have signatures of a massive tertiary at a distance of a few AU in the gravitational waveform, which may be detected at low frequencies ($f_{\rm GW}\lesssim1\rm\, Hz$) by LISA, DECIGO, TianGO, and TianQin.
    \item[(5)] There are $10^5$ undetected radio-quiet bNS systems with a massive BH tertiary in the Milky Way, and about 10 of them will have merger time of $1\rm\, Myr$ or less and are detectable by LISA/TianQin when the modulation of the outer orbit is taken into account.
\end{itemize}

\section*{Data Availability}
The data underlying this article will be shared on reasonable request to the corresponding author.

\section*{acknowledgements}
WL thank Shri Kulkarni for his encouragement throughout this project and many insightful suggestions. We thank Tony Piro for suggesting looking into low-GW-frequency observations of the predicted signals. We thank Brian Metzger, Bin Liu, Imre Bartos, Kyle Kremer, Jim Fuller, Ylva Goetberg, Pawan Kumar for many useful discussions and comments. We acknowledge the detailed explanations of their earlier work on AGN-assisted mergers by Imre Bartos and Yang Yang. WL was supported by the David and Ellen Lee Fellowship at Caltech. The research of PB and CB was funded by the Gordon and Betty Moore Foundation through Grant GBMF5076.

{\small
\bibliographystyle{mnras}
\bibliography{refs}
}

\appendix
\section{Spin of the GW190814 Secondary}
\begin{figure}
\centering
\includegraphics[width=0.47\textwidth]{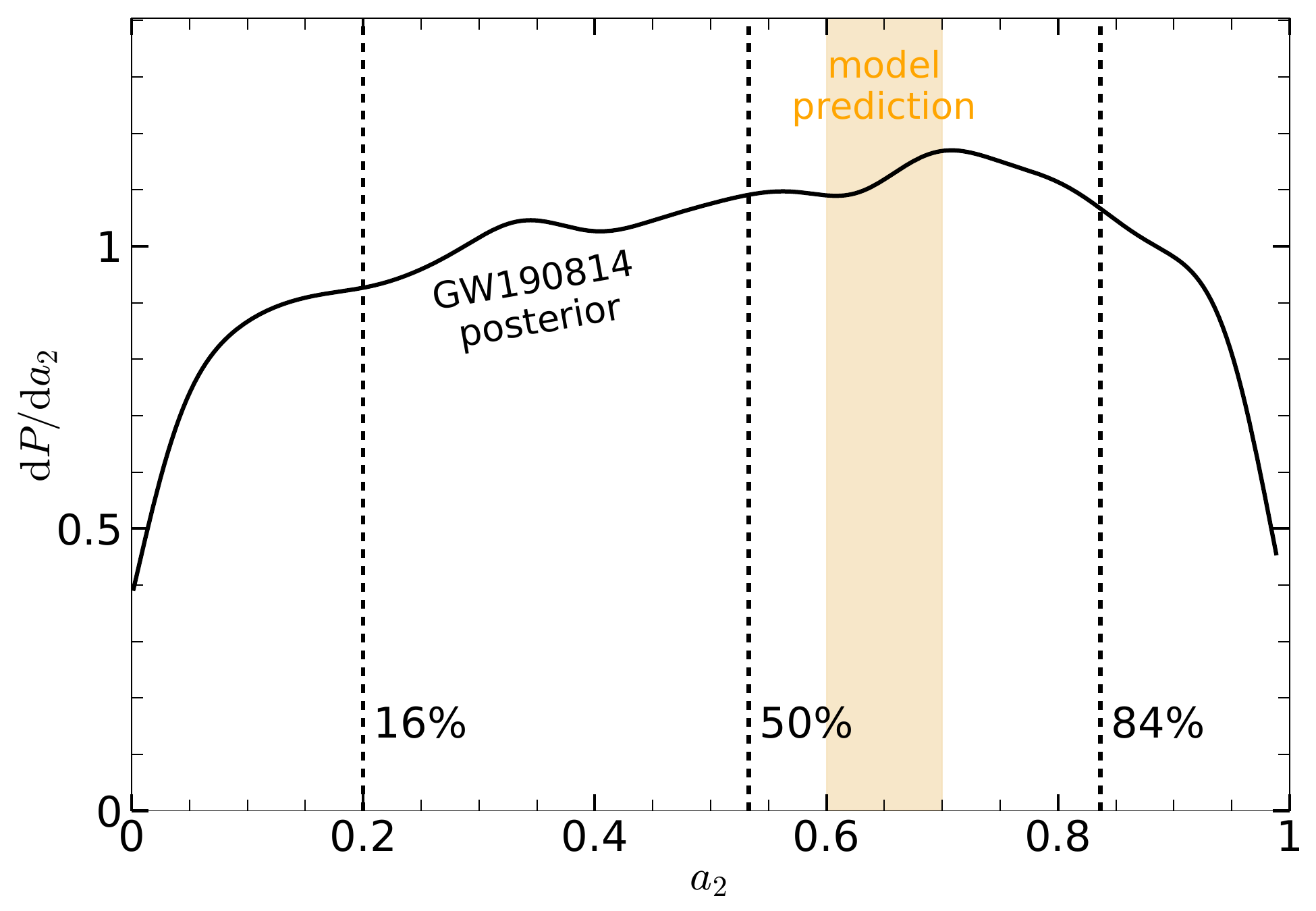}
\caption{The black-solid line shows the posterior for the magnitude of the dimensionless spin $a_2$ of the GW190814 secondary, as made publicly available by \citet{abbott20_GW190814}. Although the current constraints are not very strong ($a_2=0.53\pm0.3$ at 68\% confidence), the data slightly favors our model-predicted range of $0.6$--$0.7$, as shown in a light-orange vertical band.
}
\label{fig:secondary_spin}
\end{figure}

We show the marginalized posterior for the magnitude of the dimensionless spin $a_2$ of the GW190814 secondary in Fig. \ref{fig:secondary_spin}. Although the current constraints are not very strong, our model-predicted range of $a_2\sim 0.6$--$0.7$ is slightly favored. A more accurate spin measurement may be possible for higher signal-noise-ratio events in the future, which will provide a crucial test of our model.

\section{Markov-Chain Monte Carlo simulation}
\begin{figure}
\centering
\includegraphics[width=0.47\textwidth]{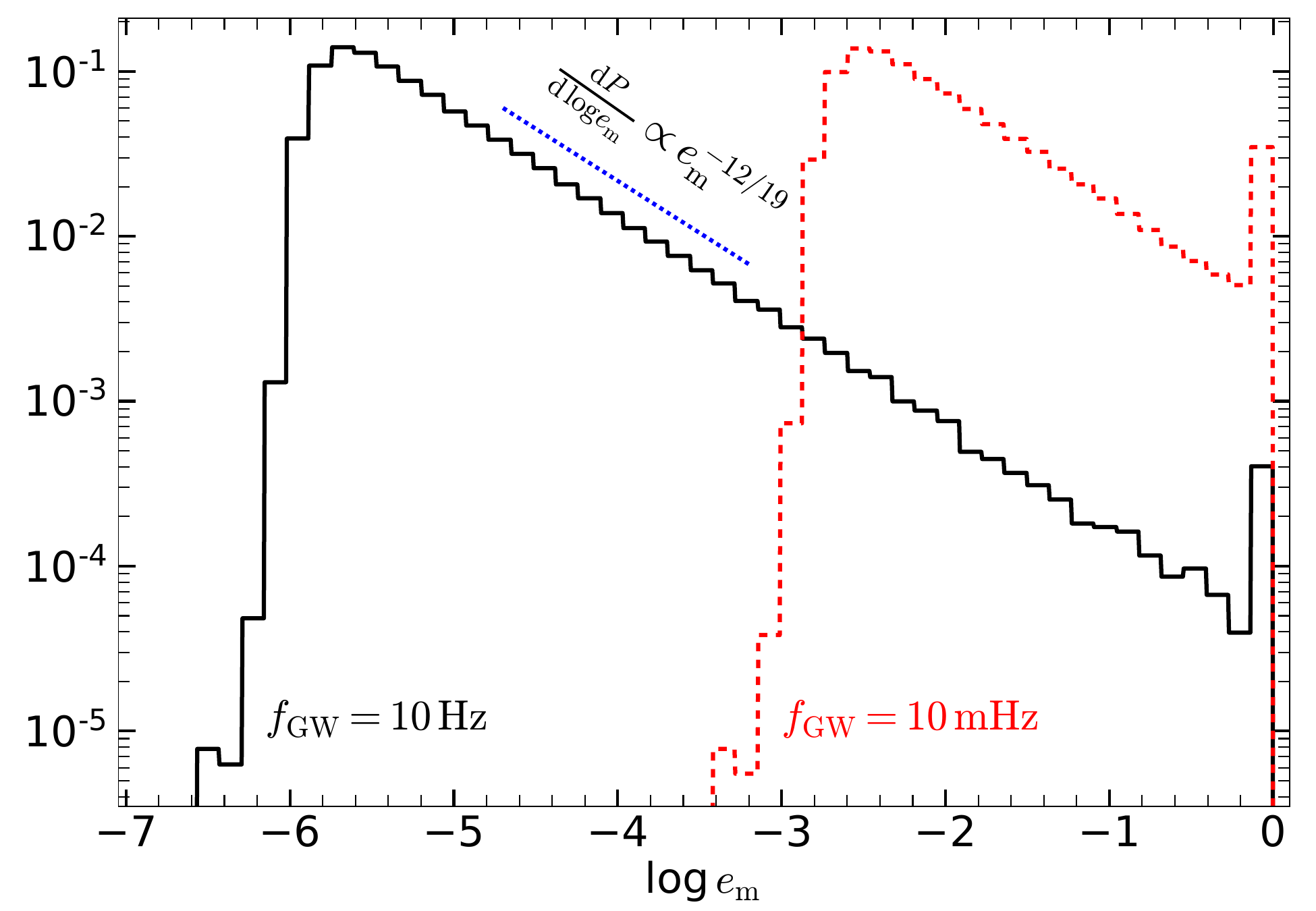}
\includegraphics[width=0.47\textwidth]{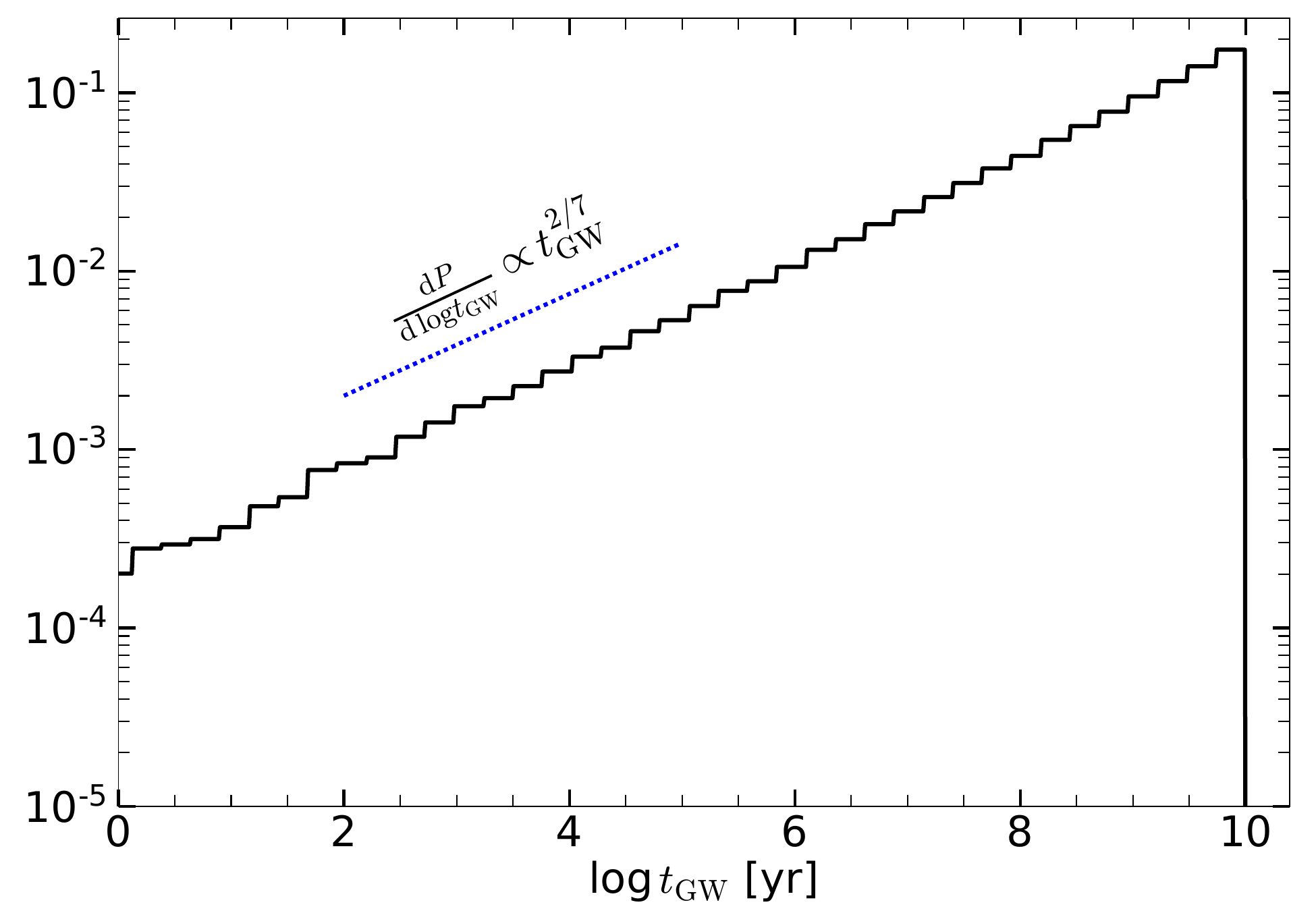}
\caption{\textbf{\textit{Upper panel}}: The PDF ($\d P/\d\, \mr{log}\,e_{\rm m}$) for the ``merger eccentricity" $e_{\rm m}$ at two different GW frequencies $f_{\rm GW}=10\rm\, mHz$ (red-dashed line, for LISA/TianQin) and $f_{\rm GW}=10\rm\, Hz$ (black-solid line, for LIGO). The peak near $e_{\rm m}\approx 1$ is due to the fact that, in a small fraction of cases, the orbital angular frequency near the pericenter is already near or above the detector frequencies ($f_{\rm GW}$) right after the bNS merger. We computed the GW-driven orbital evolution using eq. (5.11) of \citet{peters64} and then $e_{\rm m}(f_{\rm GW})$ from eq. (37) of \citet{wen03_merger_eccentricity}. For $f_{\rm GW}=10\rm\,mHz$, 90\% (or 10\%) of the cases have $e_{\rm m}\gtrsim 2\times 10^{-3}$ (or $e_{\rm m}\gtrsim 9\times10^{-2}$).
\textbf{\textit{Lower panel}}: The PDF ($\d P/\d\, \mr{log}\,t_{\rm GW}[\rm yr]$) for the GW merger time right after the capture of the remnant BH. The maximum merger time considered is 10 Gyr. Only $0.4\%$ of the cases have $t_{\rm GW}<100\rm\, yr$.
}
\label{fig:lgem_lgtgw}
\end{figure}

\begin{figure*}
\centering
\includegraphics[width=0.95\textwidth]{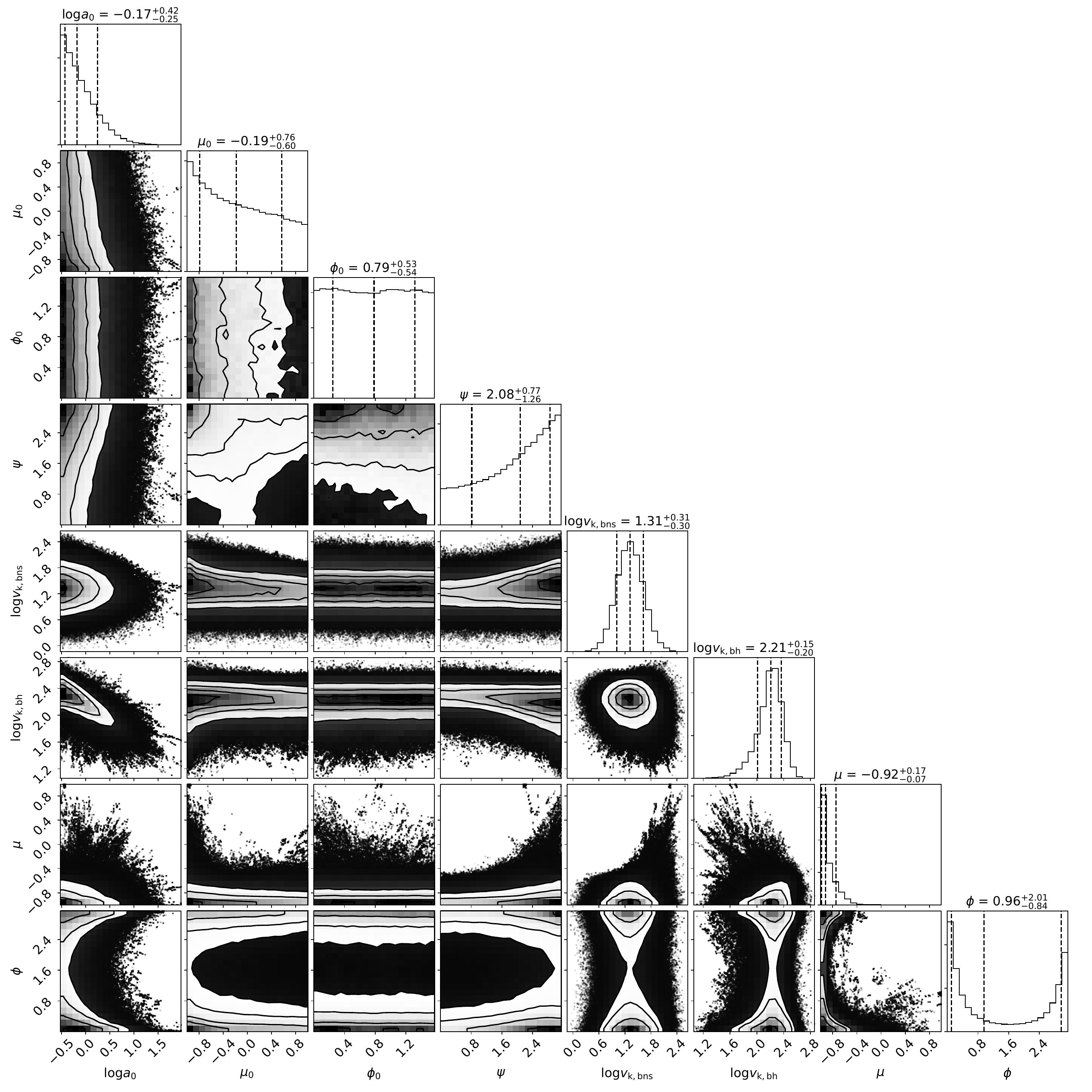}
\caption{The posterior distribution for the parameters that give rise to 2nd-generation bBH merger within a Hubble time, including initial outer SMA $a_0 [\rm AU]$, kick on the bNS system $(v_{\rm k, bns} [\mr{km/s}], \mu_0\equiv\cos\theta_0, \phi_0 \mr{[rad]})$, orbital phase angle $\psi \mr{[rad]}$ when bNS merger occurs, and kick on the remnant BH $(v_{\rm k, bh} [\mr{km/s}], \mu\equiv\cos\theta, \phi [\mr{rad}])$. We adopt flat priors in $-0.5<\mr{log}a_0[\mr{AU}]<2$ (corresponding to $p=0$ in eq. \ref{eq:SMA_distribution}), $-1<\mu_0<1$, $0<\phi_0<\pi/2$ (from symmetry), $-1<\mu<1$, $0<\phi<\pi$ (from symmetry). The priors for the two kick amplitudes $v_{\rm k,bns}$ and $v_{\rm k,bh}$ are log-normal distribution with standard deviation of 0.3 dex in log space, centered at $\bar{v}_{\rm k,bns}=20\rm\, km/s$ and $\bar{v}_{\rm k,bh}=100\rm\, km/s$, respectively. The prior for orbital phase angle $\psi$ is given by eq. (\ref{eq:prior}), which corresponds to a uniform distribution in time. We obtain the posterior from the Bayesian theorem with likelihood function equal to 1 if the merger time $t_{\rm GW}<10\rm\, Gyr$ and 0 otherwise. We fix the two BH masses as $M=20$ and $m=2.6\msun$. This figure was generated with the $\mathtt{corner}$ Python package \citep{corner}.
}
\label{fig:posterior}
\end{figure*}

We carry out Markov-Chain Monte Carlo (MCMC) simulations\footnote{We used the $\mathtt{emcee}$ Python package \citep{emcee}} to obtain the posterior distribution for all 8 parameters in our model that give rise to 2nd-generation bBH merger within 10 Gyr. From the MCMC samples, it is straightforward to calculate the marginalized distribution of any relevant quantity. In Fig. \ref{fig:lgem_lgtgw}, we show the distributions of the eccentricities when the peak GW frequency reaches $10\rm\, mHz$ and $10\rm\, Hz$, as well as the distribution of GW merger time right after the capture of the remnant BH from the bNS merger. The full MCMC posterior is shown in Fig. \ref{fig:posterior} (see the caption for the parameter used for the case).

\section{Effects of Natal Kicks on the first-born neutron star and the tertiary black hole}
Here, we discuss how the results may be affected by the kicks associated with the first-born NS and the tertiary BH, which have been ignored in our calculation.

The first-born NS was likely from a regular supernova which generated a large kick speed of $\sim$300$\rm\, km/s$ \citep{hobbs05}. Provided that the inner binary stays bound, the first-born NS is tied with the other inner-binary member which was still on the main-sequence at this time. The remaining main-sequence star might have gained some mass from Roche-lobe overflow before the supernova. If the masses of the remaining inner binary members are $15\msun$ and $1.4\msun$ (as representative values), the center of mass of the inner binary will gain a kick velocity of $1.4/(15 + 1.4)\times 300\rm\, km/s \sim 25\, km/s$. This is sufficient to dissociate the triple if the outer SMA is wider than $\sim$100$\,\rm AU$. If the outer SMA is in the range 10 to 100 AU, then the kick will lead to an appreciable amount of eccentricity.

The natal kick on the tertiary BH (which was born the very first) is highly uncertain. Using the ``momentum conserving" prescription where the linear kick momentum on a typical NS is applied to the $23\msun$ BH, one may roughly estimate the kick to be $1.4/23\times 300\rm\, km/s\sim 18\rm\, km/s$.

Thus, the two kicks we have ignored have similar effects on the outer orbit as the kick due to the second-born NS in the $v_{\rm k,bns} = 20\rm\, km/s$ case we have considered. The only difference is that after either the BH kick or the first-NS kick, additional stellar evolution might change the inner and outer orbits (e.g., the inner orbit will likely shrink due to common envelope evolution). The interplay between the (highly uncertain) evolution of massive stars and the orbits is beyond the scope of the current work \citep[see][for relevant discussions]{toonen16_triple}.

All three natal kicks will only be important if the outer SMA is wider than a few tens of AU. In this paper, we mainly focus on the triple systems with much tighter outer orbits (with $a_0 \lesssim\,$a few$\rm\,AU$, see Fig. \ref{fig:fmerger1D}). In fact, our scenario is only consistent with observations if a large fraction, $f_{\rm triple}\gtrsim 10\%$, of bNS mergers occur in such tight triples. We conclude that our results are not strongly affected by the ignorance of the natal kicks on the first-born NS and the tertiary BH. This can also be seen in Table \ref{tab:probability}, where we compare the results between $v_{\rm k,bns}=20\rm\, km/s$ (``strong kick") and $10\rm\, km/s$ (``weak kick"), and the difference in the final merger fraction is small.

\label{lastpage}
\end{document}